\documentclass[twocolumn,twoside]{article}

\usepackage{a4}
\usepackage{amssymb}
\usepackage{amsmath}
\usepackage[numbers,sort&compress]{natbib}
\usepackage{graphicx}

\newcommand{\beq}{\begin{equation}}
\newcommand{\eeq}{\end{equation}}
\newcommand{\bea}{\begin{eqnarray}}
\newcommand{\eea}{\end{eqnarray}}

\newcommand{\kB}{k_\mathrm{B}}
\newcommand{\aB}{a_\mathrm{B}}
\newcommand{\am}{a_{m}}
\newcommand{\dec}{\textrm{.}}
\newcommand{\dd}{\mathrm{d}}
\newcommand{\EF}{\epsilon_\mathrm{F}}
\newcommand{\etal}{et al.}
\newcommand{\gcc}{g~cm$^{-3}$}
\newcommand{\HT}{Hulse\,--\,Taylor}
\newcommand{\mion}{m_i}
\newcommand{\rhob}{\rho_b}
\newcommand{\omc}{\omega_c}
\newcommand{\omci}{\omega_{ci}}
\newcommand{\Tb}{T_{b}}
\newcommand{\Teff}{T_\mathrm{eff}}
\newcommand{\TF}{T_{F}}
\newcommand{\Ts}{T_{s}}

\begin{document}

\title{{\bf The physics of neutron stars}\\
{\normalsize\sc in memory of vitaly lazarevich ginzburg}}

\date{{\normalsize\textit{Ioffe Physical-Technical Institute,
Politekhnicheskaya 26, Saint Petersburg 194021,
Russia}}\\[1ex]{\small\textit{Usp.\ Fiz.\ Nauk} \textbf{180}, 
1279--1304 (2010)
[in Russian]\\
English translation: \textit{Physics -- Uspekhi}, \textbf{53}, 1235--1256}
\\{\small Translated from Russian by Yu V Morozov,
edited by A M Semikhatov and by the author}
}

\author{Alexander Y. Potekhin}

\maketitle

\begin{abstract}
Topical problems in the physics of and basic facts
about neutron stars are reviewed. 
The observational manifestations of neutron stars, their
core and envelope structure, magnetic fields, 
thermal evolution, and masses and radii are briefly
discussed, along with the underlying microphysics.
\end{abstract}

\setcounter{secnumdepth}{3}
\setcounter{tocdepth}{2}

\tableofcontents

\section{Introduction}
\label{sect:intro}

Neutron stars, the most compact stars in the Universe, were
given this name because their 
interior is largely composed of neutrons.
A neutron star of the typical
   mass $M\sim 1$\,--\,$2\, M_\odot$, where
$M_\odot=2\times10^{33}$~g is the solar mass,
has the radius $R\approx10-14$ km. The mass density
$\rho$ in such star is $\sim10^{15}$ \gcc,
or roughly 3 times \emph{normal nuclear density}
(the typical density
   of a heavy atomic nucleus)
$\rho_0=2\dec8\times10^{14}$~\gcc. 
The density $\rho$ in the center of a
   neutron star can be an order of magnitude higher than $\rho_0$. 
Such matter cannot be obtained under laboratory conditions,
and its properties and even composition remain to be
   clarified. There are a variety of theoretical models to describe
   neutron star matter, 
and a choice in favor of one of them in the
   near future will be possible only after an analysis and
   interpretation of relevant observational data using these
   models. Neutron stars exhibit a variety of unique properties
   (that are discussed below) 
and produce many visible manifestations that can be used to verify theoretical models of
   extreme states of matter \cite{Fortov}. 
Conversely, the progress 
in theoretical physics studying matter under
   extreme conditions creates prerequisites
for the construction
   of correct models of neutron stars and adequate 
   interpretation of their observations.
   
Neutron stars are not the sole objects in whose depth
   matter is compressed to high densities inaccessible in laboratory.
Other representatives of the class of \emph{compact stars}
are white dwarfs and hypothetical quark stars \cite{Glendenning}.
While the size of a neutron star mainly depends
on the balance between gravity force and
   degenerate neutron pressure, white dwarfs resist gravitational
   squeezing due to the electron 
   degeneracy pressure, and quark or strange stars resist it due
to the pressure of matter composed of
   quarks not combined into hadrons.
Neutron stars are much
   more compact than white dwarfs. White dwarfs, regarded as a
   specific class of stars since the 1910s, with the mass
$M\sim M_\odot$ have the radius
$R\sim10^4$ km, which is comparable to Earth's radius but
   almost 1000 times greater than the radius of a neutron star 
\cite{Shklovsky-book}. 
Therefore, the matter density in their interiors is less than 
one-thousandth of 
$\rho_0$. 
On the other hand, according to theoretical models, quark stars
at $M\sim M_\odot$ 
may be even more compact than neutron stars. 
But unlike neutron stars, quark stars have not been yet observed,
and their very existence is questioned.

Vitaly Lazarevich Ginzburg was among the pioneers of neutron star
   theoretical research. He predicted certain important features
   of these objects before they were discovered by radio
   astronomers in 1967 and greatly contributed to the 
   interpretation of observational data in the subsequent period. A
   few of his papers concerning these issues were published in
   1964. In Refs.~\cite{Ginzburg64,GinzburgOzernoy}
 (the latter in co-authorship with
   L M Ozernoi), he described changes in the stellar magnetic
   field during collapse (catastrophic compression) and
   obtained the value $B\sim10^{12}$~G,
   accurate to an order of
   magnitude for typical magnetic induction of a neutron star
   with a mass $M\sim M_\odot$, 
formed in the collapse. Moreover,
   Ginzburg derived expressions for the magnetic dipole field
   and the field uniform at infinity taking account of the 
   space-time curvature near the collapsed star, in accordance with the
   general relativity (GR).
Today, these expressions are
   widely used to study magnetic neutron stars. In the same
   work, he predicted the existence of a neutron star 
   magnetosphere in which relativistic charged particles emit 
   electromagnetic waves in the radio to X-ray frequency range, and
   demonstrated the influence of magnetic pressure and 
   magnetohydrodynamic instability and the possibility of detachment
   of the current-carrying envelope from the collapsing star    
 \cite{GinzburgOzernoy}. Subsequent theoretical and observational studies confirmed
the importance of these problems for the neutron star physics.
In a one-and-a-half-page note 
\cite{GinzburgKirzhnits}, Ginzburg and Kirzhnits
formulated a number of important propositions concerning
neutron superfluidity in the interior of neutron stars
(apparently independently of the earlier note by Migdal 
\cite{Migdal59}), 
the formation of Feynman--Onsager vortices, a critical
superfluidity temperature ($T_c\lesssim10^{10}$~K)
 and its dependence on
the density ($\rho\sim10^{13}$\,--\,$10^{15}$ \gcc), 
and the influence of
neutron superfluidity on heat capacity and therefore on the
thermal evolution of a neutron star. These inferences, fully
confirmed in later research, were further developed in Ginzburg's
review
\cite{Ginzburg69sup}, 
 where he considered, inter alia, the superfluidity
of neutrons and the superconductivity of a proton admixture
to the neutron fluid in the core of a neutron star. In 
Ref.~\cite{GinzburgSyr},
Ginzburg and Syrovatskii put forward the correct hypothesis of
magnetic bremsstrahlung radiation from the source of 
X-rays in the Crab Nebula that turned out to be a plerion (a
pulsar wind nebula) surrounding the neutron star and of its
origin from the envelope stripped off in collapse. In 1968,
Ginzburg
and co-workers greatly contributed to the elucidation of the
nature of \emph{radio pulsars} -- cosmic sources of periodic radio
pulses: they first developed the model of oscillating white
dwarfs \cite{GinzburgKirzhniz,GinzburgZZ68}, and later
 the models of rotating neutron
stars with strong magnetic fields 
\cite{GinzburgZ69,GinzburgZZ69,GinzburgZ75}; in these studies, they
discussed putative mechanisms of pulsar radiation 
\cite{GinzburgZ70a,GinzburgZ70b}. Specific works were devoted
to the role of pulsars in generation of cosmic rays
\cite{Ginzburg71hi}  and the estimation of the
work function needed to eject ions from the pulsar surface
into the magnetosphere \cite{GinzburgUsov}. In 1971, Ginzburg published a
comprehensive review \cite{Ginzburg71},
focused on the analysis of
theoretical concepts of the neutron star physics and the
physical nature of pulsars formulated by that time. The
review contained a number of important original ideas, such
as the estimation of typical neutron star magnetic fields
 $B\sim10^{12}$~G, where he has pointed out that lower values to
$B\sim10^{8}$~G and higher values to
$B\sim10^{13}$\,--\,$10^{15}$~G are also possible.
These estimates were brilliantly confirmed by later studies, 
which showed that
a maximum in the magnetic field distribution of
radio pulsars lies around $B\sim10^{12}$~G
\cite{ATNF}; millisecond pulsars
discovered in the 1980s have $B\sim10^{8}$\,--\,$10^{10}$~G
\cite{Bisno06}; and fields of magnetars discovered in the 1990s
reach up to
$B\sim10^{14}$\,--\,$10^{15}$~G \cite{Mereghetti08}. 
We must note that many of Ginzburg's other findings have wide
application in neutron star research. 
Besides the famous studies on superfluidity and 
superconductivity, 
one should mention in this respect his
investigations into the distribution of electromagnetic
waves in the magnetized plasma, summarized in the
comprehensive monograph
\cite{Ginzburg}.

Neutron stars are still insufficiently well known
   and remain puzzling
objects despite their extensive studies in many research centers
during the last 40 years. The aim of this review, from the
perspective of modern astrophysics, is to highlight the main
features of neutron stars making them unique cosmic bodies.
It does not pretend to be comprehensive, bearing in mind the
thousands of publications devoted to neutron stars, and is
designed first and foremost to present the author's personal
view of the problem. Fundamentals of neutron star physics
are expounded in the excellent textbook by Shapiro and
Teukolsky \cite{ShapiroTeukolski}. More detailed descriptions
 of developments in
selected branches of neutron star astrophysics can be found in
monographs (such as
\cite{Glendenning,NSB1}) and specialized reviews 
(published, among others, in \textit{Physics--Uspekhi} -- i.e.,
\cite{Yak-super,Beskin99,Bisno06}). 

In Sect.~\ref{sect:basics} we give
general information on the physical
properties of neutron stars and related physical and 
astrophysical problems; the history of relevant research is also
outlined.
Section \ref{sect:obs}
equally concisely illustrates the ``many faces'' of neutron
stars as viewed by a terrestrial observer. The following
sections are of a less general character, each being concerned
with a specific aspect of astrophysics. The list of these aspects
is far
from exhaustive. For example, we will not consider
the physics of the pulsar
magnetosphere and the mechanisms underlying generation
of its radiation, which are
dealt with in the voluminous literature (see, e.g.,
\cite{GinzburgZ75,Beskin99,Malov04,Michel}). The
same refers to nucleon superfluidity discussed in 
comprehensive reviews (e.g., \cite{Yak-super}).
Only in passing are mentioned the neutrino emission mechanisms,
exhaustively described in the reviews of D G Yakovlev with coworkers
\cite{Yak-super,Yak-neu}. 
The list of references on the problems concerned in this review
is neither exhaustive,
nor even representative, which could not be otherwise, keeping in view
the format of the article. The author apologizes to those
researchers whose important contributions to the physics of
neutron stars are not cited in this publication. A more
complete (although still nonexhaustive)
bibliography can be found in monograph \cite{NSB1}.

\section{Basic facts about neutron stars\label{sect:basics}}

\subsection{Neutron stars as relativistic objects\label{sect:GR}}

Of great importance for neutron stars, unlike ordinary stars,
are the GR effects \cite{MisnerTW}. 
The
structure of nonrotating stars is described by the 
nonrelativistic equation of hydrostatic equilibrium for a spherically
symmetric body in GR, that is the 
\emph{Tolman--Oppenheimer--Volkoff} (TOV) equation 
\cite{Tolman39,Oppenheimer39}.
It also gives a very good
approximation for rotating stars, except those with 
millisecond rotation periods. The minimal possible period is
 $\sim0\dec5$ ms,
 but that observed to date is almost thrice as large,
1.396 ms 
\cite{Hessels_ea06}, characteristic of the ``slow rotation regime'' in
which the effects of rotation can be taken into account in
terms of the perturbation theory [\citenum{NSB1}, Ch.~6]. The corrections
introduced by the magnetic field are negligibly small for the
large-scale structure of a neutron star (at least for
$B\lesssim 10^{16}$~G). The effects of the known magnetic fields
$B<10^{15}$~G can be important in stellar envelopes,
as we will discuss in Sect.~\ref{sect:mag}. Solution of the TOV equation for a
given equation of state of neutron star matter yields a family
of stellar structure models, whose parameter is $\rho_c$,
the density in the center of the star. The stability condition requiring that 
 $M(\rho_c)$ be an
increasing function is satisfied within a certain range of stellar
masses and radii; the maximum mass $M_\mathrm{max}$, compatible with
the modern theory, is approximately 
$M_\mathrm{max}\approx1\dec5$\,--\,$2\dec5\,M_\odot$,
depending on
the equation of state being used, while the minimal possible
mass of a neutron star is 
$M_\mathrm{min}\sim0\dec1\,M_\odot$. The significance of
the GR effects for a concrete star is
determined by the compactness parameter
\beq
 x_g=r_g/R,
 \quad \mbox{where~}
 r_g=2GM/c^2 \approx 2\dec95\,M/M_\odot \textrm{ km} 
\label{r_g}
\eeq
is the Schwarzschild radius, $G$ is the gravitational constant, 
and $c$ is the speed of light. Gravity at the stellar surface 
is determined by the equality  
\beq 
g=\frac{GM}{R^{2}\,\sqrt{1-x_g}}
\approx \frac{1\dec328\times10^{14}}{\sqrt{1-x_g}}\,\frac{M/M_\odot}{
\,R_6^{2}}\textrm{ cm s}^{-2}, 
\eeq 
where $R_6\equiv R/(10^6\mbox{~cm})$. 

The \textit{canonical} 
neutron star is traditionally a star with
$M=1\dec4\,M_\odot$ and $R=10$ km ($g=2\dec425\times
10^{14}\textrm{ cm s}^{-2}$).
   Note that the best and most detailed equations of state
   available to date predict a slightly lower compactness:
    $R\approx12$ km at
$M=1\dec4\,M_\odot$ (see Sect.~\ref{sect:coreobs}). Substituting
   these estimates in (\ref{r_g}), we see that the effects of general
   relativity for a typical neutron star amount to tens of
   percent. This has two important consequences: first, the
   quantitative theory of neutron stars must be wholly 
   relativistic; second, observations of neutron stars open up unique
   opportunities for measuring the effects of general relativity
   and verification of their prediction.

The near-surface photon frequency (denoted by 
$\omega_0$) in a locally inertial reference frame undergoes a
redshift to $\omega_\infty$ according to
\beq
 z_g \equiv \omega_0/\omega_\infty -1  = (1-x_g)^{-1/2} -1.
\label{z_g}
\eeq
Therefore, the thermal radiation spectrum of a star with an
   effective temperature $T_\mathrm{eff}$, measured by a distant observer,
   is displaced toward longer 
   wavelengths and corresponds to a lower effective temperature
$T_\mathrm{eff}^\infty = T_\mathrm{eff} \, \sqrt{1-x_g}$. 

Along with the radius $R$, determined by the
equatorial length $2\pi R$ in the locally inertial
reference frame, one often introduces
the \emph{apparent radius} for
a distant observer: $R_\infty = R \,(1+z_g)$. 
In particular, for the canonical neutron star
we have $R_\infty=13$ km, and for a more realistic model of the same
   mass, $R_\infty\sim15$
km. The radius  $R$ of a neutron star decreases
   as its mass increases, but the growth of
   $z_g$ with the reduction
   of the radius and the increase in the mass leads to the appearance
   of a
   minimum in the dependence
$R_\infty(M)$. One can show that the
   apparent stellar radius cannot be smaller than
$R_\infty^\mathrm{min}=7\dec66\,(M/M_\odot)$~km
\cite{NSB1}.
The overall apparent photon
   luminosity 
$L_\gamma^\infty\propto R_\infty^2 (T_\mathrm{eff}^\infty)^4$
is related to the luminosity in the local stellar reference frame
as
$L_\gamma^\infty = (1-x_g)\, L_\gamma$.
The expressions for $R_\infty$ and $L_\gamma^\infty$ 
are in excellent agreement with the
   notion of light bending and time dilation in the vicinity of a
   massive body. The light bending enables a distant
   observer to ``look behind'' the horizon
   of a neutron star.
   For example, the
   observer can simultaneously see both polar caps of a star
   having a dipole magnetic field at a proper dipole inclination
   angle to the line of sight. This effect actually occurs in
   observation of pulsars. Naturally, such effects must be taken
   into account when comparing theoretical models and
   observations.

According to GR, 
 a rotating star having a shape other
   than the ellipsoid of revolution can emit gravitational
   waves. Shape distortions may be caused by star oscillations
   and other factors. It has been speculated \cite{BonazzolaG} that 
   gravitational waves emitted by rapidly rotating neutron stars can be
   recorded by modern gravitational antennas. However, these
   antennas appear more suitable for recording gravitational
   waves from merging neutron stars \cite{Brag00,PoPro07}.

While gravitational waves have not yet been detected by
   ground-based antennas, they have already been registered in
   observations of ``space antennas'' -- double neutron stars.
   Two stars orbiting a common center of masses are known to
   emit gravitational waves. The first pulsar rotating in an orbit
   together with another neutron star was discovered by Hulse
      and Taylor in 1974 (Nobel Prize of 1993). It remained the sole
   known object of this kind for over 20 years. At least 9 such
      systems have been described to date. The most remarkable of
   them is the \emph{double pulsar} system J0737--3039, a binary in
   which \emph{both} neutron stars are seen as radio pulsars 
\cite{Bisno06,KramerStairs}. 

The known binary systems of neutron stars have compact
   orbits and short periods of revolution. The orbital period of
   the \HT\ pulsar is less than 8 hours, and the large
   semiaxis of the orbit is about two million kilometers, or
   almost two orders of magnitude smaller than the distance
   between the Sun and Earth. Gravitational radiation is so
   strong that the loss of energy it carries away results in a
   significant decrease in both the orbit size and the orbital
   period. The measured decrease in the orbital period of the
\HT\ pulsar is consistent with that predicted by GR
   within a measurement error to a few tenths of a percent.

Another GR effect is the periastron shift
   or relativistic precession of the orbit that is orders of
 magnitude greater than Mercury's perihelion shift (to be precise,
   those  $7\dec5$\% of its perihelion shift, 
   namely $0\dec43''$ per year, that cannot be
      accounted for by the influence of other objects of the Solar
system are explained in the GR framework). For
example, for the \HT\ pulsar
the relativistic periastron shift is
$4\dec22^\circ$ per year,
and for the double pulsar it is $16\dec9^\circ$ per year. 

 The third measured effect is geodesic precession of a
   rotating body that moves in an orbit, a precession analogous
   to the spin-orbital interaction in atomic physics. The
   measurement of geodesic precession made it possible to
   reconstruct the time dependence of the direction of the
    \HT\ pulsar magnetic axis. It turned out that its
   directivity pattern would no longer intersect the line of sight
   of an Earth-based observer around 2025, and the pulsar
   would become invisible for two centuries 
\cite{Istomin91,Kramer98}.
One cannot exclude that the companion star will become visible.

The double pulsar proved an even better laboratory than
   the \HT\ pulsar for the verification of the effects of
   general relativity. First, registration of radio pulses from both
   neutron stars of the binary system allows directly measuring
   the radial velocities from the Doppler shift and geodesic
   precession of either star from the altered pulse shape.
   Second, the line of sight of a terrestrial observer lies virtually
   in the plane of the double pulsar orbit (at the inclination to the
   normal
$\approx89^\circ$). This permitted for the first time to reliably
   measure the so-called Shapiro delay parameters (two 
   parameters characterizing the time delay of an electromagnetic
   wave passing a star.\footnote{In addition,
   the proximity of the orbit plane to the line of sight allowed
 one to observe modulation of pulsed emission from one pulsar passing through
   the atmosphere of the other, which provided supplementary
   information on their magnetic fields and magnetospheres.} 
   Five of the seven independent 
   post-Kepler parameters characterizing the effects of general
   relativity were measured for the double pulsar. Any two of
   them uniquely define masses of both pulsars $M_A$ and
$M_B$,
   while the measurement of the remaining ones may be
   regarded as verification of GR. 
   HR has brilliantly passed this
   test: any two of the measured parameters gave the same values 
$M_A=1\dec337\,M_\odot$, $M_B=1\dec249\,M_\odot$ 
within measurement errors $<0\dec001\,M_\odot$ \cite{KramerStairs}.

 One more unique relativistic object discovered in 2005 is
   the pulsar PSR J1903+0327, having the rotation period
$2\dec15$ ms and moving in an inclined highly elliptical orbit
   (eccentricity $e=0\dec44$ and inclination $78^\circ$)
   in a pair with a main-sequence star (an ordinary star of a mass
$M\approx M_\odot$, identified in the infrared \cite{Champion}).
Observations with the Arecibo radio telescope yielded three
   post-Kepler parameters: both Shapiro delay parameters and
   orbital precession. The estimate
$M=1\dec67\pm0\dec01\,M_\odot$
was
   obtained for the pulsar mass under the assumption that the
   orbital precession is due to the effects of GR alone
    \cite{Freire}. It is
   the largest mass of neutron stars measured thus far. We note,
   however, that the influence of nonrelativistic effects, such as
   tides at the companion star caused by gravitational attraction
   of the pulsar, on the orbital precession cannot be totally
   excluded. With this uncertainty, a more conservative estimate
   is
$M=1\dec67\pm0\dec11\,M_\odot$ \cite{Freire}.
\emph{(See the Note added in proof).}

\subsection{\label{sect:str}The biggest enigmas of 
the neutron star structure}

\begin{figure}
\begin{center}
\includegraphics[width=.6\columnwidth]{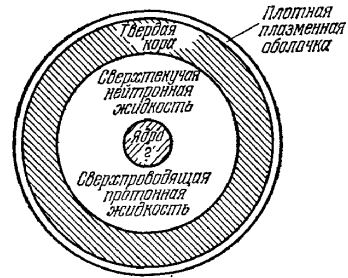}
\end{center}
\caption{Schematic of the neutron star structure, from
the paper \cite{Ginzburg71} by V L Ginzburg.
The labels read (from the center outwards): Core,
Superfluid neutron liquid, Superconducting proton liquid,
Rigid crust, Dense plasma envelope.
\label{fig:nsgross}}
\end{figure}

Two main qualitatively different regions, the core and the
   envelope, are distinguished in a neutron star. The core is in
   turn subdivided into the outer and inner core, and the
   envelope into the solid crust and the liquid ocean. Such a
   division into four essentially different regions was proposed
   in the review by Ginzburg of 1971 \cite{Ginzburg71} 
   (Fig.~\ref{fig:nsgross}). \emph{The outer core} 
of a neutron star
   is usually several kilometers thick, and its matter density is
    $0\dec5 \rho_0
\lesssim \rho \lesssim 2\rho_0$. This matter, accounting for the largest
   fraction of the stellar mass, has well-known qualitative
   characteristics (see, e.g.,
[\citenum{NSB1}, Ch.~5 and 6]). It is a neutron superfluid
   with an admixture of the superconducting proton component
   (see Fig.~\ref{fig:nsgross}), as well as electrons and
$\mu^-$-mesons (muons), all these
   constituents being strongly degenerate. 
   \emph{The inner core} occupies the central part with
$\rho\gtrsim2\rho_0$ and has a radius to
   several kilometers. It can be present in rather massive neutron
   stars, $M\gtrsim1\dec4$\,--\,$1\dec5\,M_\odot$ (in
   less massive neutron stars, density does not reach $2\rho_0$). 
Neither the composition nor the
   properties of matter in the inner core are known because the
   results of their calculation strongly depend on the theoretical
   description of collective fundamental interactions. From this
   standpoint, studies of neutron stars are important not only
   for astrophysics but also for nuclear and elementary particle
   physics. The available theoretical models presume the
   following hypothetical options:

\begin{list}{1.}{
\setlength{\itemsep}{0ex}
\setlength{\parsep}{0pt}
\setlength{\parskip}{0pt}
\setlength{\topsep}{0pt}
\setlength{\partopsep}{0pt}
}
\item[1.] hyperonization of matter -- the appearance of various
   hyperons (first of all, $\Lambda$- and $\Sigma^-$-hyperons);
\item[2.] pion condensation -- formation of a Bose condensate
   from collective interactions with the properties of $\pi$-mesons;
\item[3.] kaon condensation -- formation of a similar 
     condensate from
$K$-mesons;
\item[4.] deconfinement -- phase transition to quark matter.
\end{list}
The last three variants, unlike the first one, are not feasible
   for all modern theoretical models of matter with supranuclear
   density; therefore, they are frequently called 
\emph{exotic}
[\citenum{NSB1}, Ch.~7].

According to current concepts, the core of a neutron star
   contains superfluid baryonic matter. Superfluidity reduces
   the heat capacity of this matter and the neutrino reaction rate.
   However, superfluidity may be responsible for an additional
   neutrino emission due to Cooper pairing of nucleons at a
   certain cooling stage in those parts of the star where the
   temperature decreases below critical values. These effects and
   their influence on the cooling rate of a neutron star are
   reviewed in \cite{Yak-super}.

In the stellar \emph{envelopes} the matter is not so extraordinary:
the atomic nuclei are present separately there. 
Nevertheless, this matter also occurs
   under extreme (from the standpoint of terrestrial physics)
   conditions that cannot be reproduced in the laboratory. This
   makes such matter a very interesting subject of plasma
   physics research \cite{Fortov}. Equally important is the fact that an
   adequate theoretical description of stellar envelopes is
   indispensable for the correct interpretation of characteristics
   of the electromagnetic radiation coming from the star, i.e., for
   the study of its core by means of comparison of theoretical
   models and astronomical observations.

\subsection{\label{sect:nslife}The birth, 
life, and death of a neutron star}

A neutron star is a possible end product of a main-sequence
   star (``normal'' star) \cite{Shklovsky-book}. Neutron stars are believed to be
   formed in type-I supernovae explosions 
\cite{Shklovsky-SNe,ImshennikNadyozhin,Imshennik95,Arnett,WoosleyJanka}.
An 
   explosion occurs after a precursor to a supernova has burned out its
   nuclear ``fuel'': first hydrogen, then helium produced from
   hydrogen, and finally heavier chemical elements, including
   oxygen and magnesium. The end product of subsequent
   nuclear transformations is isotopes of iron-group elements
accumulated in the center of the star. The pressure of the
   electron Fermi gas is the sole factor that prevents collapse of
   such an iron-nickel core to its center under the force of
   gravity. But as soon as a few days after oxygen burning, the
   mass of the iron core increases above the Chandrasekhar limit
   equal to
$1\dec44\,M_\odot$, which is the maximum mass whose
   gravitational compression is still counteracted by the 
   pressure of degenerate electrons. Then
    \textit{gravitational collapse}, i.e.,a catastrophic breakdown of the stellar core, occurs. It is
   accompanied by the liberation of an enormous gravitational
   energy ($\gtrsim10^{53}$~erg) and a shock wave that strips off the outer
   envelopes of the giant star at a speed amounting to 10\%
 of the
   speed of light, while the inner part of the star continues to
   contract at approximately the same rate. The atomic nuclei
   fuse into a single giant nucleus. If its mass surpasses the
    \emph{Oppenheimer\,--\,Volkoff limit}, 
    that is the maximum mass that the
   pressure of degenerate neutrons and other hadrons is able to
   support against gravitational compression
   ($\approx2$\,--\,$3\,M_\odot$),according to modern theoretical models), the compression
   cannot be stopped and the star collapses to form a black hole.
   It is believed that the collapse resulting in a black hole may be
   responsible for the flare from a \emph{hypernova},
    hundreds of times
   brighter than a supernova; it may be a source of mysterious
   gamma-ray bursts coming from remote galaxies 
   \cite{Paczynski98,Postnov99}. If the
   mass remains below the Oppenheimer\,--\,Volkoff limit,, a
   neutron star is born whose gravitational squeezing is
   prevented by the pressure of nuclear matter. In this case,
   about 1\%  of the released energy transforms into the kinetic
   energy of the envelopes flying apart, which later give rise to a
   nebula (\emph{supernova remnant}), 
   and only $0\dec01$\% ($\sim10^{49}$~erg)
   into electromagnetic radiation, which 
   nevertheless may overshine the luminosity of the entire galaxy
   and is seen as a supernova.

Not every star completes its evolution as a supernova (not
   to mention a hypernova); only massive stars with
    $M\gtrsim8\,M_\odot$are destined to have such a fate. A less massive star at the end
   of its lifetime goes through a giant phase, gradually throwing
   off the outer envelopes, and its central part shrinks into a
   white dwarf.

A newborn neutron star has the temperature above
$10^{10}$\,--\,$10^{11}$~K
; thereafter, it cools down (rather fast initially,
   but slower and slower afterwards), releasing the energy in the
   form of neutrino emission from its depth and electromagnetic
   radiation from the surface. But the evolution of a neutron star
   is not reduced to cooling alone. Many neutron stars have
   strong magnetic fields that also evolve through changes in
   strength and configuration. A rotating neutron star having a
   strong magnetic field is surrounded by an extended plasma
   magnetosphere formed due to the knockout of charged
   particles from the surface by the rotation-induced electric
   field, thermal emission, and the birth of electron-positron
   pairs upon collisions of charged particles of the 
   magnetosphere with one another and with photons. Given a
   sufficiently fast rotation of a star, its magnetosphere 
   undergoes collective acceleration of the constituent particles in the
   parts where plasma density is too low to screen the strong
   electric field induced by rotation. Such processes generate
   coherent directed radio-frequency emission due to which the
   neutron star can be seen as a radio pulsar if it rotates such that
   its directivity pattern intersects observer's line of sight. The
   rotational energy is gradually depleted and the particles born
   in the magnetosphere have a charge whose sign is such that
   the induced electric field makes them propagate toward the
 star; they accelerate along the magnetic force lines, hit the star
   surface near its magnetic poles, and heat these regions. A
   similar process of heating magnetic poles occurs in the case of
 \emph{accretion} (infall of matter) onto a star, e.g., as it passes
   through dense interstellar clouds or as the matter outflows
   from the companion star in a binary system. The hot polar
   caps emit much more intense X-rays than the remaining
   surface; as a result, such neutron stars look like \emph{X-ray
   pulsars}. Pulsed X-ray radiation is also observed from thermonuclear
explosions of accreted matter at the surface of a
   rotating neutron star (see, e.g., review
\cite{StrohmayerBildsten}).

Cooling, changes in the magnetic field, or a slowdown of rotation
may cause \emph{starquakes}, associated with variations of the
crustal shape, phase transformations in the core, and interaction
between the normal and superfluid components of the core and the
crust  
\cite{Ruderman69,BP71,Blaes-ea89,HDP90,Alpar98,Franco-ea}. 
Starquakes are accompanied by liberation of the thermal energy
and sharp changes in the character of rotation
\cite{Reisen-roto}.  Moreover, the matter falling onto the star
during accretion undergoes nuclear transformations at the surface
and, due to its weight, causes additional transformations in the
depth of the envelope that alter the nuclear composition and
liberate energy \cite{BisnoChech,HZ90,HZ03}. In other words,
neutron stars not only cool down but also are heated from the
inside. 

A single neutron star eventually exhausts its supply of
   thermal and magnetic energy and fades away. A star has more
   promising prospects for the future if it is a member of a binary
   system. For example, if the companion star overfills its Roche
   lobe (the region in which matter is gravitationally coupled to
   the companion), this matter accretes onto the neutron star so
   intensely as to make it a bright source of X-ray radiation by
   virtue of the released gravitational and thermonuclear energy.
   In this case, the inflowing matter forms an \emph{accretion disk}
   around the neutron star, which also radiates X-ray emission,
   and this luminosity changes with time, e.g., as a result of disk
   precession or variations of the accretion rate. The character of
   accretion strongly depends on neutron star magnetization
   and the rotation period \cite{Lipunov}. If the mass of the accreted matter
   surpasses a critical threshold, the neutron star collapses into a
   black hole.

If the companion of a neutron star is a compact object,
   then the radius of their mutual orbit may be small enough to
   enable gravitational waves emitted by such a system to
   appreciably influence its evolution. The orbital radius of the
   compact binary system decreases as gravitational radiation
   continues until the two companions merge together, giving
   rise to a black hole with the release of enormous gravitational
   energy comparable to the stellar rest energy $\sim
Mc^2\sim10^{54}$ erg
   in the form of neutrino and gravitational radiation (this will
   happen to the \HT\ pulsar and the double
   pulsar in some 300 and 85 million years, respectively). 

One can say that \emph{the three main driving forces} 
of the evolution of a
   neutron star, responsible for its observational manifestations,
   are \emph{rotation, accretion, and magnetic field}.

\subsection{\label{sect:nshist}The formation of neutron star concepts}

Baade and Zwicky \cite{BaadeZwicky} theoretically predicted neutron stars
   as a probable result of supernova explosions less than 2 years
   after the discovery of the neutron \cite{Chadwick32}.They also put forward
   the hypothesis (now universally accepted) that supernovae
   are important sources of galactic cosmic rays and coined the
   term ``supernova'' itself to differentiate between these 
   unusually bright objects formed in a gravitational collapse giving
   rise to a neutron star from more numerous nova stars originating, as
   known today, from thermonuclear burning of the accreted
   matter at the surface of white dwarfs. The popular belief
that Landau predicted
   neutron stars in 1932 
\cite{ShapiroTeukolski}, based on recollections by Leon Rosenfeld
  \cite{Rosenfeld74}, is not accurate:the meeting
   of Landau with N Bohr and L Rosenfeld occurred in 1931,
   before the discovery of neutrons. Nevertheless, it is true that
   Landau already foresaw the existence of neutron stars at that
   time and suggested a hypothesis according to which stars
   with a mass greater than $1\dec5\,M_\odot$ 
   have a region in their interior
   where "the density of matter
   ``becomes so great that atomic nuclei come in close contact,
forming one gigantic nucleus''
\cite{Landau32}.

Not a single neutron star was observed for 43 years after
   Baade and Zwicky's prediction. But theorists continued to
   work. In 1938, Zwicky \cite{Zwicky38} estimated
the maximum binding
   energy of a neutron star and the gravitational red shift of
   photons emitted from its surface. A few months later, Tolman
    \cite{Tolman39}  and Oppenheimer and Volkoff
     \cite{Oppenheimer39} derived the 
   aforementioned TOV equation; moreover, the latter authors
   computed the limiting mass of a neutron star, $M_\mathrm{max}$, 
   although it proved underestimated because they ignored
   baryon-baryon interactions. Equations of state of nuclear
   matter began to be extensively studied in the 1950s. In 1959,
   Cameron
\cite{Cameron59} obtained the first realistic estimate of
$M_\mathrm{max}\approx2\,M_\odot$. He was the first to show that the core of a
   neutron star may contain hyperons. In the same year, Migdal
   \cite{Migdal59},
based on the concept of superfluidity in atomic nuclei
   proposed by A Bohr, Mottelson, and Pines \cite{BMP58}, predicted
   superfluidity of neutron star matter. In 1960, Ambartsumyan
   and Saakyan \cite{AS60} 
constructed the equation of state of superdense
   matter by taking electrons, muons, and all hadrons known at
   that time into consideration. They hypothesized that the core
   of a neutron star consists of two components, the outer
   composed of nucleons and the inner containing hyperons.
   The following year, Zeldovich
\cite{Zeld61} derived an extremely stiff
   equation of state of a neutron star in which the speed of
   sound tends to the speed of light as the density
   increases. Finally, in the 1960s, the first estimates of neutrino
   emission from the interior of a neutron star
    \cite{ChiuSalpeter64,BahcallWolf} and its cooling
\cite{ChiuSalpeter64,Stabler60,Morton64,BahcallWolfb,tc66} were reported; in addition, the presence of a
   strong magnetic field was predicted 
\cite{Ginzburg64,GinzburgOzernoy} and the deceleration
   of rotation of a magnetized neutron star due to 
   magnetodipole radiation calculated \cite{Pacini67}.

The first simplest models of neutron star cooling already
   demonstrated that the surface temperature of a typical
   neutron star might be as high as hundreds of thousands or
   millions of degrees, meaning that the star emits thermal
   radiation largely in the form of soft X-rays unable to
   penetrate Earth's atmosphere. The progress in astronautics
   and the advent of X-ray astronomy in the early 1960s
   \cite{Giacconi62} gave hope that such radiation would be detected in outer
   space. However, it took almost 30 years to reliably identify
   thermal X-ray components in the spectra of neutron stars
   by the X-ray telescope on board the ROSAT
satellite, witch
   produced images with a resolution of a few angular
   seconds 
\cite{BeckerTrumper}. 

Other means were proposed to search for neutron stars.
   Zeldovich and Guseinov \cite{zg66} suggested that they could be
   detected in binary systems with optical stars from the Doppler
   shifts of optical spectral lines. Kardashev 
\cite{Kardashev} and Pacini \cite{Pacini67} put forward the
correct hypothesis that the rotational energy of a
   neutron star was transferred via the magnetic field to the
   surrounding nebula formed in the collapse at stellar birth.
   These authors regarded the Crab Nebula as such a candidate
   object. However, neutron stars unexpectedly manifested
 themselves as radio pulsars.

As known, the first pulsar was discovered by radio
   astronomers at Cambridge in 1967 \cite{Hewish68}
    (a retrospective
   study of the archives of the Cambridge group gave evidence of
   radio pulses from pulsars dating back to 1962--1965 
   \cite{Hewish75}).
   Antony Hewish, who headed the research team, was awarded
   the Nobel Prize in physics in 1974 for this achievement. A 
   correct explanation
   of these observations soon after their publication was
   proposed by Thomas Gold in the paper entitled
    ``Rotating neutron stars as the origin of the pulsating
radio sources'' \cite{Gold68}. It is
   less widely known that Shklovsky \cite{Shklovsky67} arrived at the
   conclusion, based on analysis of X-ray and optical 
   observations, that radiation from Scorpio X-1 (the first X-ray source
   discovered outside the Solar System \cite{Giacconi62})) originated from the
   accretion of matter onto a neutron star from its companion.
   Unfortunately, this conclusion (later fully confirmed
    \cite{deFreitas77}) was accepted too sceptically at that time
\cite{CameronScoX1}.

The discovery of pulsars gave powerful impetus to the
   development of theoretical and observational studies of
   neutron stars. With over one thousand publications devoted
   to these celestial bodies appearing annually, a new class of
   astronomical objects containing neutron stars is discovered
   once every few years. For example, X-ray pulsars were
   described in 1971, bursters (sources of X-ray bursts) in 1975,
   soft gamma repeaters (SGR5) in 1979, millisecond pulsars in
   1982, radio-silent neutron stars in 1996, anomalous X-ray
   pulsars (AXPs) in 1998, and rapid (rotating) radio transients
   (RRATs) in 2006. Clearly, it is impossible to cover all these
   developments in a single review. We try instead to depict the
   current situation in certain important branches of the theory,
   although we start with observable manifestations of neutron
   stars.

\section{\label{sect:obs}Observational manifestations of neutron stars}

Radiation from neutron stars is observed in all ranges
of the electromagnetic spectrum.
As well as 40 years ago, most of them (about 1900 as of 2010
 \cite{ATNF}) are seen as radio pulsars. Some 150 of
   the known neutron stars are members of binary systems with
   accretion and manifest themselves largely in the form of 
   X-ray radiation from the accretion disk or flares produced by
   explosive thermonuclear burning in the star outer layers.
   Certain such systems make up X-ray transients in which
   periods of active accretion (usually as long as several days or
   weeks) alternate with longer periods of quiescence (months or
   sometimes years) during which X-ray radiation from the hot
   star surface is recorded. In addition, over one hundred
   isolated neutron stars are known to emit X-ray radiation.

\subsection{\label{sect:INS}Cooling neutron stars}

A large fraction of emission 
from isolated neutron stars and X-ray transients in quiescence
   appears to originate at their surface. To interpret
   this radiation, it is very important to know the properties of
   envelopes contributing to the spectrum formation. 
   Conversely, comparison of predictions and observations may be used
   to deduce these properties and to verify theoretical models of
   dense magnetized plasma. Moreover, investigating the 
   properties of the envelopes provide knowledge of the parameters
   of a star as a whole and of the observational constraints on
   such models.

For each theoretical model of neutron stars, a
\emph{cooling
   curve} describes the dependence of the overall photon
   luminosity
$L_\gamma^\infty$ in the frame of a distant observer
on time $t$ elapsed after the birth of the star (see review
 \cite{YakovlevPethick} and the references
   therein).

There are not many neutron stars in whose spectrum the
 cooling-related thermal component can be distinguished from
   the emission produced by processes other than surface
   heating, e.g., those proceeding in the pulsar magnetosphere,
   pulsar nebula, and accretion disk. Fortunately, there are
   exceptions \cite{Zavlin09}, such as relatively young 
   ($t\lesssim10^5$years)
   pulsars J1119--6127,  B1706--44, and Vela, whose spectra are
   readily divisible into thermal and nonthermal components, and
   medium-aged ($t\sim10^6$ years) pulsars
B0656+14,  B1055$-$52, and Geminga. 
The spectra of the latter three objects, dubbed
``three musketeers''
\cite{BeckerTrumper,Zavlin09}, are fairly well
   described by the three-component model (power-law 
   spectrum of magnetospheric origin, thermal spectrum of the hot
   polar caps, and thermal spectrum of the remaining surface).
   Even more important is the discovery of radio-silent neutron
   stars \cite{Walter96},with purely thermal spectra. These are 
   central compact
   objects (CCOs) in supernova remnants \cite{DeLuca08} 
and X-ray ``dim'' isolated neutron stars (XDINS) \cite{Haberl07}. 
Observations indicate that CCO may have magnetic fields
 $B\sim10^{10}$\,--\,$10^{11}$~G (slightly weaker than in the majority of
   normal pulsars but stronger than in millisecond pulsars) 
   \cite{HalpernGotthelf}, and XDINSs may have magnetic fields 
$B\gtrsim10^{13}$~G (somewhat stronger than ordinary)
\cite{Mereghetti08}. As many as 7 XDINSs are constantly known during the last
decade, and astrophysicists call them
the ``Magnificent Seven''
\cite{Haberl07,Zavlin09,PopovProkhorov}. 
By the way, confirmed CCOs count also seven, but three more objects
are candidates waiting to be included in the list
 \cite{HalpernGotthelf}. The spectra of at least 5
   radio-quiet neutron stars exhibit wide absorption lines for
   which no fully satisfactory theoretical explanation
    has been proposed thus far.
   Certain authors hypothesize that they can be attributed to
   ionic cyclotron harmonics in a strong magnetic field, but
   rigorous quantum mechanical calculations  \cite{PC03,P10}
   have proved that such harmonics in neutron star atmospheres are too weak
   to be observed.

Besides cooling processes, heating processes of different
   natures occur in neutron stars. Sometimes they compete with
   the heat delivered to the stellar surface from the core and must
   be taken into consideration. Such stars include:
\vspace*{-1ex}
\begin{itemize}\itemsep=0pt
\item[---]
 old neutron stars ($t\gtrsim10^6$ years) for which the cooling
   curves would go down to the low temperature
   region ($T_\mathrm{eff}\lesssim10^5$~K) when disregarding heating;
\item[---]
magnetars, which are relatively young
($t\lesssim10^4$ years) neutron stars with superstrong
 ($B\gtrsim10^{14}$~G) magnetic fields 
   manifested as AXPs and SGRs
\cite{Mereghetti08,PopovProkhorov}. The strong X-ray luminosity
   of magnetars cannot be explained by the ``standard cooling
   curve'' \cite{YakovlevPethick}.
   Thompson \cite{Thompson01}
   suggested that it should be ascribed
   to heating due to dissipation of a superstrong magnetic field.
   Recent studies \cite{UrpinKonenkov08,Pons_MG,Kaminker_ea09}
provided some support to this hypothesis.

\end{itemize}

The discussion of neutron star cooling is to be continued
   in Section \ref{sect:cool}. For now, we emphasize that the processes of
   cooling, heating, and heat transfer turn the surface of a
   neutron star into a source of thermal radiation with a
   spectral maximum in the soft X-ray region.

\subsection{Pulsars}
\label{sect:PSR}

Pulsed radiation related to the proper rotation of neutron
   stars contains important additional information
\cite{ShapiroTeukolski,ManchesterTaylor,Malov04}. For example, simultaneous measurement of emission at
   several radio frequencies allows determining the 
   \emph{dispersion measure}
   from the phase shift, which in turn permits roughly
   estimating the distance to a pulsar. Measurements of the
   pulsation amplitude of the thermal component in the
   spectrum characterize the nonuniform temperature distribution over the surface. The period of pulsations 
    $P$ and its time derivative $\dot{P}$ for isolated (nonaccreting) pulsars give an idea of
   the star magnetic field (of its dipole constituent, to be precise)
   and age:
\beq
   B\sim10^{19\dec5}\sqrt{\dot{P}P/\mbox{1~s}}\textrm{~~G},
    \quad
     t\sim t_\mathrm{PSR}\equiv 0\dec5P/\dot{P},
\label{PPdot} 
\eeq 
where $t_\mathrm{PSR}$ is the so called \emph{characteristic age}
of a pulsar.
Such estimate makes no sense for accreting pulsars because
   their period of rotation may depend on the interaction
   between the magnetic field and the accretion disk. For
   example, the rotation can be accelerated by virtue of the
   transfer of angular momentum from matter falling onto the
   pulsar; then $\dot{P}<0$.

We note that the age of a neutron star can be deduced
   from the age of the remnant of the supernova hosting this
   star. As a rule, the age of the remnant is inferred with an error
    $\sim10$\% from the rate at which the envelopes fly apart from
   each other. Certainly, the remnant itself must be accessible to
   observation, which is rarely the case (as a rule, only at
$t\lesssim10^4$ years). When both the characteristic age and the age of the
   remnant are known, they are consistent with each other to the
   order of magnitude, but their numerical difference may be as
   great as 2--3-fold. This means that none of the age estimation
   methods is entirely reliable. The exceptions are 5 supernovae
   whose flares are documented in historical chronicles
   \cite{GreenStephenson}.

Simultaneous measurement of the age and magnetic field
   permits imposing limitations on the decay rate of the stellar
   magnetic field. The relevant theoretical estimates are 
   significantly different, depending on the field configuration inside
   the crust and the 
   core, which in turn depends on the model and on the hypothesis
    of the nature of the field; they also depend on
   the electric conductivity included in a given model, which is in
   turn a function of the poorly known chemical composition of
   the envelopes, the microscopic structure of the crust, the
   content of admixtures, and the defects of the crystal lattice.

îX-ray radiation from pulsars, similarly to radio emission,
   carry important information. Generally speaking, the X-ray
   spectrum of pulsars contains both thermal and nonthermal
   components. The latter is produced in the magnetosphere
   either by synchrotron radiation or inverse Compton 
   scattering of charged particles accelerated to relativistic energies by
   magnetospheric electromagnetic fields. This component is
   usually described by a power-law spectrum. The thermal
   component is divided into ``hard'' and ``soft'' constituents. The
   former is supposed to be a result of radiation from the \emph{polar
   caps} heated to millions of degrees, where the magnetic field
   does not substantially deviate from the normal to the surface.
   In the dipole field model, the radius of these regions is
   estimated as
\beq
R_\mathrm{cap}\approx\left(\frac{2\pi R^3}{cP}\right)^{1/2}
\approx 0\dec145 R_6^{3/2}\,(P/\textrm{1 s})^{-1/2}\textrm{~km}.
\label{polarcap}
\eeq
The soft constituent corresponds to radiation from the
      remaining, cooler surface that may be due to the heat coming
      up from the stellar core or inner crust.

\subsubsection{Ordinary pulsars}

Ordinary
pulsars are isolated pulsars with
      periods from tens of milliseconds to several seconds. Their
      characteristic magnetic field given by formula
(\ref{PPdot}) varies from a
      few gigagauss to
$10^{14}$~G with typical values $B\sim10^{11}-10^{13}$~G, and the
      characteristic age is from several centuries to $10^{10}$
years with typical values $t_\mathrm{PSR}\sim10^5-10^8$ years
\cite{ATNF}. The X-ray spectrum of thermal
      radiation from certain normal pulsars has been measured,
      which allows the cooling theory methods to be applied to their
      study. Unlike the case of neutron stars lacking pulsation, the
      estimates of $t_\mathrm{PSR}$ and
$B$ make the class of thermal structure
models more definite and thereby restrict the scattering of
possible cooling curves.

\subsubsection{Millisecond pulsars}

Millisecond pulsars have magnetic
   fields $B\sim10^8-10^{10}$~G
and are from tens of millions to
   hundreds of billions of years old,
   with typical values
   $t_\mathrm{PSR}\sim10^9-10^{10}$ years (with the exception of
PSR J0537$-$6910 with abnormally high $\dot{P}$)
\cite{ATNF}. The relatively weak magnetic field and the
   short period of millisecond pulsars may result from the pulsar
   having passed through the stage of accretion in the course of
   its evolution, which reduced the magnetic field and increased
   the angular momentum due to the interaction between the
   accreting matter and the magnetic field \cite{Lipunov,Bisno06}. Certain
   isolated millisecond pulsars emitting in the X-ray range
   show a thermal constituent in their spectra produced by
   radiation from the hot polar caps \cite{Zavlin07}.
It is convenient to
   write formula (\ref{polarcap})
for millisecond pulsars in the form
 $R_\mathrm{cap}\approx R_6^{3/2}\,(P/\textrm{21
ms})^{-1/2}\textrm{~km},$ which readily shows that the hot
   region covers a large surface area of the pulsar.

\subsubsection{Anomalous pulsars}

Many neutron stars
   manifest themselves through pulsed radiation in the X-ray
   part of the spectrum. They are referred to as X-ray pulsars.
   Some of them are located in binary stellar systems. Evidently,
  those radio pulsars
   that exhibit thermal radiation from polar caps are also
    X-ray pulsars.

Unlike these ``normal'' X-ray pulsars, AXPs have an
   unusually long period, 
$P\approx6-12$~s and high X-ray luminosity $\sim10^{33}-10^{35}$
erg s$^{-1}$, being in the same time isolated
\cite{Mereghetti08,MalovMachabeli09}. Their magnetic fields and characteristic ages
   estimated from (\ref{PPdot}),
suggest that these objects may be
   magnetars. An alternative explanation of their properties is
   based on the assumption that they are neutron stars with
``normal'' magnetic fields $B\sim10^{12}$~G that slowly accrete
   matter from the disk remaining after the supernova 
   explosion \cite{ErtanAXP}. In other words, 
   the nature of these objects remains
   obscure.
   
\subsection{Neutron stars in binary systems}
\label{sect:BIN}

A neutron star in a binary system is paired with another
   neutron star, a white dwarf, or an ordinary (nondegenerate)
   star. Binary systems containing a neutron star and a
   companion black hole are unknown. Measuring the 
   parameters of the binary orbit provides additional characteristics
   of the neutron star, e.g., its mass.
     
     The infall of matter onto a neutron star is accompanied by
   the liberation of energy, which turns the system into a source
   of bright X-ray radiation. Such systems are categorized into
   markedly different subclasses: low-mass X-ray binary 
   systems with a dwarf (either white or red) of mass 
   $\lesssim2M_\odot$ as the
   companion, and relatively short-lived massive systems in
   which the mass of the companion star is several or tens of
   times greater than $M_\odot$and accretion of matter onto the
   neutron star is extremely intense.

X-ray binaries may be sources of regular (periodic) and
   irregular radiation, and are subdivided into permanent and
   temporary (transient). Emission from some of them is
   modulated by neutron star rotation, others are sources of
   quasiperiodic oscillations (QPOs), bursters (neutron stars whose
   surface from time to time undergoes explosive 
   thermonuclear burning of the accreted matter), and so on.

The QPOs,
first observed in 1985~\cite{vanderklisetal85},
 occur in X-ray binary systems containing compact objects,
   such as neutron stars (typically within low-massive X-ray
   binaries), white dwarfs, and black holes. There are a variety of
   hypotheses on the nature of the QPOs 
   (see \cite{vanderklis00,Kluz07,Tagger07}).
They seem to originate in the accretion disk. According
   to some hypotheses, they are related to the Kepler frequency
   of the innermost stable orbit permitted by GR, a resonance in the
   disk, or a combination of these frequencies with the rotation
   frequency of the compact object. Given that one of these
   hypotheses is true, the QPOs in low-massive X-ray binary systems may become a tool for
   determining the parameters of neutron stars.

X-ray luminosity of type-I bursts in bursters may reach the
    \emph{Eddington limit}
$
L_\mathrm{Edd}
  \approx
1\dec3\times10^{38}\,(M/M_\odot) 
$
erg s$^{-1}$,
at which
   radiation pressure on the plasma due to the Thomson
   scattering exceeds the force of gravity. Such flares are of
   special interest in that simulation of their spectra and
   intensity permits estimating the parameters of a neutron star
    \cite{ShaposhnikovTitarchuk,Steiner}.

The spectra of certain soft X-ray transients during ``quiescent''
   periods exhibit the thermal radiation component of the
   neutron star located in the system This allows comparing
   the cooling curves with observations, as in the case of isolated
   neutron stars, with the sole difference that the energy release
   due to accretion must be taken into account. On the one hand,
   this introduces an uncertainty in the model, but on the other
   hand it permits verifying theoretical considerations 
   concerning accretion onto the neutron star and thermonuclear
   transformations of matter in its envelopes.

Of special interest are quasipermanent transients, i.e.,
   those whose active and quiescent periods last a few years or
   longer. According to the model proposed in \cite{Brown_BR98}, the
   thermal radiation in the periods of quiescence is due to the crust
   cooling after deep heating by accretion. Such cooling is
   independent of the details of the star structure and 
   composition and therefore its analysis directly yields information on
   the physics of envelopes. Three sources of this kind are
   known: KS
1731$-$260, MXB 1659$-$29, and AX J1754.2$-$2754
\cite{BrownCumming09}. The crust of a neutron star gets heated during a
   long period of activity and relaxes to the quasiequilibrium
   state in the quiescence period that follows, meaning that evolution
   of the thermal spectrum contains information about the
   properties of the crust. Therefore, analysis of the
   thermal luminosity of such an object and its time dependence
   provides information about heat capacity and thermal
   conductivity of the crust in the period of activity preceding
   relaxation and about the equilibrium luminosity at rest. Such
   data may in turn be used to obtain characteristics of the star
   as a whole \cite{Shternin07}.

Estimation of neutron star masses from the Kepler
   parameters of X-ray binary systems is not yet a reliable
   method because of theoretical uncertainties, such as those
   related to the transfer of angular momentum via accretion.
   The most accurate estimates are obtained for binary systems
   of two neutron stars due not only to the absence of accretion
   but also to the marked GR effects, whose measurement allows determining the
   complete set of orbital parameters. Sufficiently accurate mass
   estimates (with an error $<0\dec2\,M_\odot$)
 are also available for
   several binary systems containing a white dwarf as the
   companion and for the system mentioned in Section \ref{sect:GR}, in
   which the companion is a main-sequence star. These estimates
   lie in the range $1\dec1\,M_\odot<M<1\dec7\,M_\odot$.

\section{The core of a neutral star
   and supranuclear density matter\label{sect:core}}

In this section, we focus on the equation of state of the
   neutron star core, disregarding details of the microscopic
   theory and nonstationary processes. We note that the Fermi
energy of all particles essential for this equation is many
   orders of magnitude higher than the kinetic thermal energy.
   Therefore, a good approximation is given by the equation of
   state of cold nuclear matter in which the dependence of the
   pressure on density and temperature, $P(\rho,T)$ 
, is replaced
   by a one-parametric dependence $P(\rho)$ at $T\to0$.

\subsection{The outer core\label{sect:outer_core}}

Nucleons in the outer core of a neutron star form a strongly
   interacting Fermi liquid, whereas leptons make up an almost
   ideal Fermi gas. Therefore, the energy density  $\mathcal{E}$ can be
   represented as the sum of three terms, 
\begin{equation}
\mathcal{E}(n_n,n_p,n_e,n_\mu)=\mathcal{E}_N(n_n,n_p)
+ \mathcal{E}_e(n_e) + \mathcal{E}_\mu(n_\mu),
\label{eoscore-E.npemu}
\end{equation}
where $n_e$, $n_\mu$, $n_n$, and $n_p$ are
concentrations of electrons, $\mu^-$-mesons, neutrons, and protons. The equations of state and
   concentration of particles are determined by the energy
   density minimum at a fixed baryon volume density
$n_b=n_n+n_p$ and under the electroneutrality condition
$n_e+n_\mu=n_p$.
This implies that the relations
 $\mu_n=\mu_p+\mu_e$ and
$\mu_\mu=\mu_e$ for chemical potentials $\mu_j$ 
of particles $j=n,p,e,\mu^-$, expressing the equilibrium conditions
   with respect to the electron and muon beta-decay 
   and beta-capture reactions: 
$n\rightarrow p + e + \overline{\nu}_e$,~ $
   p+e\rightarrow n + \nu_e
$,~ $n\rightarrow p + \mu + \overline{\nu}_\mu$,
and
   $p+\mu \rightarrow n + \nu_\mu$,
  where $\nu_{e,\mu}$ and $\overline{\nu}_{e,\mu}$ are
   electron and muon neutrinos and antineutrinos. Neutron star
   matter (unlike the matter of a protoneutron star, i.e., the
 collapsed core within the first minutes after the supernova
   explosion) is transparent to neutrinos: therefore, the chemical
   potentials of neutrino and antineutrino are equal to zero.
   Electrons at the densities being considered are 
   ultrarelativistic, and therefore
$\mu_e\approx c p_{\mathrm{F}e} \approx 122\dec1\,
(n_e/0\dec05 n_0)^{1/3}~\textrm{MeV}$, 
where $p_{\mathrm{F}e}$ is the electron Fermi momentum.
and $n_0=0\dec16$ fm$^{-3}$ is the normal nuclear
number density, which corresponds to the normal nuclear
mass density $\rho_0$.
In the general case,
   muons are moderately relativistic, which dictates the use of the
   general expression 
$\mu_\mu=m_\mu c^2\sqrt{1 +
p_{\mathrm{F}\mu}^2/(m_\mu c)^2}$.
As soon
   as the equilibrium is known, the pressure can be found from
   the equation
$P=n_b^2
{{\dd}(\mathcal{E}/n_b)/ {\dd} n_b}$.

Thus, the construction of the equation of state for the
   outer core of a neutron star reduces to the search for the
   function
$\mathcal{E}_N(n_n,n_p)$. A number of ways to address this
   problem have been proposed based on a variety of 
   theoretical physics methods, viz, the Brueckner--Bethe--Goldstone
   theory, the Green's function method, variational methods,
   the relativistic mean field theory, and the density functional
   method [\citenum{NSB1},
\S\,5.9]. The Akmal--Pandharipande--Ravenhall
   (APR) model known in several variants is currently regarded
   as the most reliable one \cite{APR}. The APR model uses the
   variational principle of quantum mechanics, under which an
   energy minimum for the trial wave function is sought. This
   function is constructed by applying the linear combination of
   operators describing admissible symmetry transformations in
   the coordinate, spin, and isospin spaces to the Slater
   determinant consisting of wave functions for free nucleons.
   APR variants differ in the effective potentials of 
   nucleon-nucleon interaction used to calculate the mean energy. The
   potentials borrowed by the authors from earlier publications
   take the modern nuclear theory into account and their 
   parameters are optimized so as to most accurately reproduce the
   results of nuclear physics experiments. We note that the
   addition of an effective three-particle nucleon-nucleon
   potential to the two-particle one ensures a remarkably close
   agreement between theory and experiment.
   
The effective functional of nuclear matter energy density
   was used to construct another known equation of state, SLy
\cite{SLy}. Calculations based on the SLy equation are less detailed
   but easier to use than in the APR model. This equation is
   constructed in accordance with the same scheme as the 
   well-known FPS equation of state \cite{FPS}, which was
   especially popular in the 1990s in calculations of the
   astrophysical properties of neutron stars. The main 
   difference between SLy and FPS lies in the specification of
   parameters of the effective energy density functional 
   accounting for current experimental data. An important advantage of
   both models over many others is their applicability not only
     to the stellar core but also to the crust, which allows
   determining the position of the crust-core interface in a 
   self-consistent manner \cite{SLy-edge}.

Note by the way that there are
a convenient parametrization for the
   APR equation of state \cite{HeiselHj00} and 
   explicit fitting expressions
   for Sly \cite{HP04} for the dependence of pressure on density and the
   so-called pseudoenthalpy, a convenient parameter for 
   calculating the properties of rapidly rotating neutron stars 
   \cite{Bocquet}.

\begin{figure}
\begin{center}
\includegraphics[width=\columnwidth]{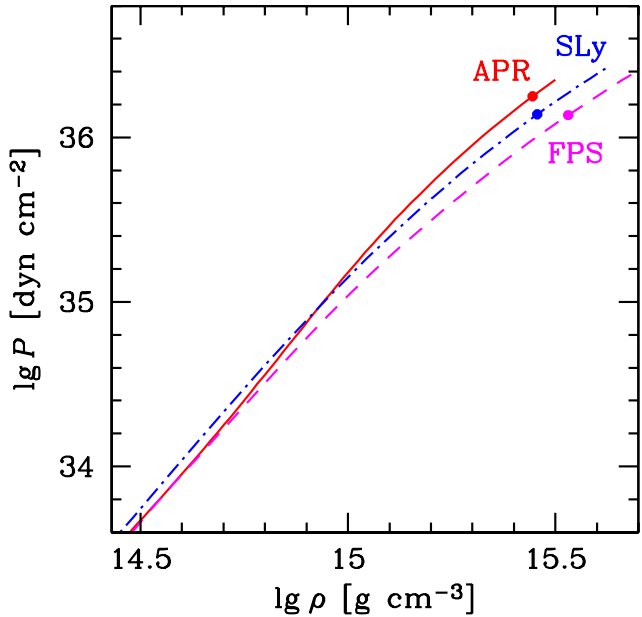}
\end{center}
\caption{FPS, Sly, and APR equations of state for the core of a neutron
   star. Bold dots on the curves correspond to the maximally possible density
   in a stationary star.
\label{fig:sly}}
\end{figure}

Figure ~\ref{fig:sly} shows $P(\rho)$ dependences for models
FPS, SLy, and APR. Comparison of FPS and Sly shows that a more
   exact account of current experimental data makes the $P(\rho)$ dependence steeper and the equation of state stiffer. Bold dots
   correspond to the density in the center of a neutron star with $M=M_\mathrm{max}$ for each of 
   these equations; the segments of the
   curves to the right of these dots cannot be realized in a static
   star.

A common drawback of the above models is the
   application of a Lorentz non-invariant theory to the 
   description of hadrons. Such a description becomes a priori incorrect
   in the central part of the core, where the speeds of nucleons on
   the Fermi surface may constitute an appreciable fraction of
   the speed of light. The same drawback is inherent in all other
   aforementioned approaches, with the exception of the
   relativistic mean field theory. This theory, suggested in the
   1950s, was especially popular in the 1970s \cite{Walecka}. It has a
   number of appealing features. Specifically, its Lorentz
invariance guarantees the fulfillment of the condition that the
   speed of sound does not exceed the speed of light, which is
   subject to violation in some other models. But the assumption
   of spatial uniformity of meson field sources underlying this
   theory is valid only if $n_b\gg100\, n_0$ \cite{NSB1}.
The matter density
   necessary for this condition to be satisfied is much higher than
   that in the interior of a neutron star. Therefore, the equations
   of state in the core of neutron stars are calculated realistically
   based on nuclear interaction models that are not Lorentz
   invariant but are still applicable to the largest part of the
   stellar core.

Along with the $P(\rho)$ dependence, it is important to know 
relative abundances of various particles. In particular, the
   dependence of the proton fraction $x_p$ in the
   neutron-proton-electron-muon ($npe\mu$) matter
on density $\rho$. The fact is that the principal
   mechanism behind neutrino energy losses in the outer core
   of a neutron star is the so-called modified Urca processes
   (in short, \emph{Murca}, after K P Levenfish) consisting of
   consecutive reactions $n+N\to p+N+e+\bar\nu_e$ and $p+N+e\to
   n+N+\nu_e$, where
$N=n$ or $N=p$ is a nucleon-mediator (``an active spectator'',
according to Chiu \& Salpeter \cite{ChiuSalpeter64}). The involvement of the 
mediator distinguishes
   the Murca processes from ordinary processes of beta-decay
   and beta-capture, referred to as direct Urca processes.\footnote{The
   term \emph{Urca process} was coined by Gamow and 
   Sch\"onberg \cite{GamowS}.
   Gamow recalled 
\cite{Gamow}: ``We called it the Urca process partially to
   commemorate the casino in which we first met and partially because the
   Urca process results in a rapid disappearance of thermal energy from the
   interior of a star similar to the rapid disappearance of money from the
   pockets of the gamblers in the Casino da Urca. Sending our article on the
   Urca process for publication in the Physical Review, I was worried that the
   editors would ask why we called the process `Urca'. After much thought I
   decided to say that this is the short for unrecordable cooling agent, but
   they never asked.''}. If $x_p\lesssim x_c$, where the value of
$x_c$ varies from $0\dec111$ to $0\dec148$ depending on the muon
abundance, then
the energy and
   momentum conservation laws cannot be fulfilled 
   simultaneously without the participation of a mediator nucleon in the
   Urca process, keeping in mind that the momenta of the
   involved strongly degenerate neutrons $n$, protons
$p$, and electrons $e$ lie near their Fermi surfaces
\cite{Haensel95}. If $x_p>x_c$, then
direct Urca processes much more powerful than Murca come
   into play. For this reason, the neutron star suffers \emph{enhanced
   cooling} if $x_p$ exceeds
$x_c$.

 In different variants of the APR model, the fraction of $x_p$ grows nonmonotonically from
 $\approx0\dec01$\,--\,$0\dec02$ at $n_b=n_0$ to
$x_p\approx0\dec16$\,--\,$0\dec18$ at $n_b>1\dec2$ fm$^{-3}$.
Therefore, if the APR
   model holds, the density of matter in the center of a massive
($M\gtrsim1\dec8\,M_\odot$) 
 neutron star is such that it switches on direct
   Urca processes and speed up its cooling. In contrast, direct
   Urca processes are impossible (for stable stars) in the Sly
   model.

\subsection{The inner core and hyperons\label{sect:inner_core}}

\begin{figure}
\begin{center}
\includegraphics[width=\columnwidth]{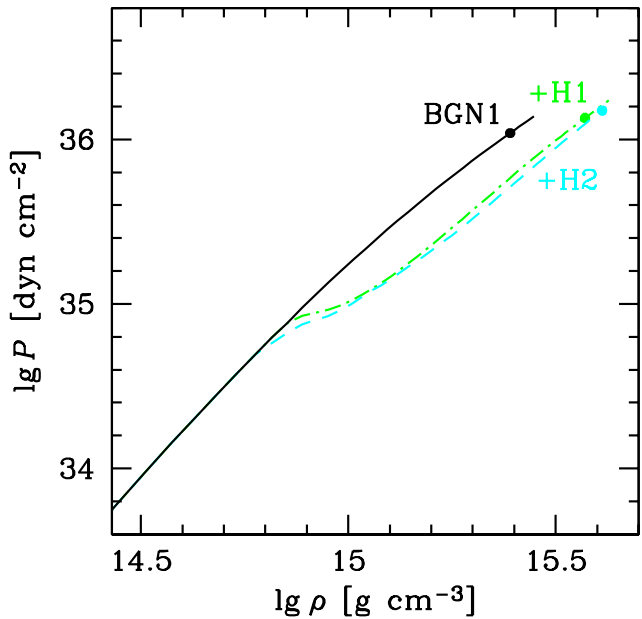}
\end{center}
\caption{Equations of state for the core of a neutron star according to the
   variants of the model proposed in \cite{BGN} with (BGN1H1 and BGN1H2)
   and without (BGNI) hyperons. Bold dots on the curves indicate the
   maximally possible density in a stationary star. The data and notations are
   borrowed from \cite{NSB1}.
\label{fig:bgn}}
\end{figure}

Strong gravitational compression in the interior of a neutron
   star is likely to provoke the conversion of nucleons into
   hyperons if such conversion may reduce the energy density
   at a given
$n_b$.  The process is mediated by the weak interaction
   with a change of strangeness (quark flavor). According to
   modern theoretical models, the conversion is possible at
$\rho\gtrsim2\rho_0$. 

The equation of state containing hyperons is calculated as
   described for the $npe\mu$ case in Sect.~\ref{sect:outer_core},
    but the equations for
   the chemical potentials are supplemented with new ones for
the equilibrium conditions with respect to the weak 
   interactions. The lightest baryons make an octet of two nucleons ($p$ 
   and $n$ with zero strangeness $S=0$), four hyperons with $S=-1$  
   ($\Lambda^0$, $\Sigma^-$, $\Sigma^0$, and
$\Sigma^+$), and two hyperons with $S=-2$ ($\Xi^0$ and $\Xi^-$). Here, they are listed in the order of increasing
   mass. Under normal conditions, hyperons decay for fractions
   of nanoseconds. But in matter composed of degenerate
   neutrons,$\mu_n$ increases with increasing density.
   When $\mu_n$ reaches a minimal chemical potential of a hyperon given by
   its mass, this hyperon becomes stable because the decay
   reaction ceases to be thermodynamically favorable.

Some clarification is needed here. Although  $\Lambda^0$ is the lightest of
   all hyperons, $\Sigma^-$ is the first to be stabilized as the density
   increases. It occurs so, because the decay of the
$\Sigma^-$ hyperon
yields an
   electron, in conformity with the equilibrium condition
    $\mu_{\Sigma^-}=\mu_n+\mu_e$. The electrons, like neutrons, are strongly
   degenerate, and their chemical potential
$\mu_e$ is equal to the
   Fermi energy; the addition of this energy to $\mu_n$
permits the
   equilibrium condition to be satisfied at a smaller density (as
   was first noticed by Salpeter in 1960 
\cite{Salpeter60}). Similarly, the
   necessity of subtracting $\mu_e$
from $\mu_n$ can make formation
of $\Sigma^+$ hyperons disadvantageous. However, electrons are
   gradually replaced by $\Sigma^-$ hyperons, and $\mu_e$ decreases as the
   density increases. Due to this, many theoretical models
   predict the appearance of $\Sigma^+$ hyperons at a sufficiently high
   density, 
$n_b\gtrsim5\,n_0$. In the general case, both electrons and
   muons are gradually substituted by negatively charged
   hyperons as the density increases. In the models predicting a
   high hyperon concentration, leptons disappear at $n_b\gtrsim1$ fm$^{-3}$
    and the so-called ``baryonic soup'' is cooked with high average
   strangeness per baryon (almost $-1$ in the central parts of
   maximum-mass stars).

The current theory lacks a rigorous description of
   nucleon-hyperon and hyperon-hyperon interactions. This
   uncertainty is aggravated by the uncertainty arising from the
   choice of the mode of description of multiparticle 
   interactions; this results in a great variety of model equations of state
for the inner core of a neutron star. Figure~\ref{fig:bgn} 
exemplifies three
   of the equations of state proposed in \cite{BGN}. 
 The solid curve
   corresponds to the so-called minimal model disregarding
   hyperons. The dashed and dashed-dotted curves correspond
   to two models with hyperons. If the density exceeds the
   threshold for the appearance of new particles, the equation
   of state is noticeably softened, as is natural when part of the
   strongly degenerate high-energy neutrons are replaced by
   slow heavy hyperons. However, the magnitude of the effect
   depends on the details of hyperon-nucleon and hyperon-hyperon interactions.

\subsection{Phase transformations and deconfinement\label{sect:deconf}}

As the density $\rho$ increases above the nuclear density $\rho_0$  matter
   may undergo phase transitions to qualitatively new states
   regarded as exotic from the standpoint of terrestrial nuclear
   physics; the very existence of these states depends on the
   concrete features of strong interactions and the quark
   structure of baryons.

\subsubsection{Meson condensation}

It has been known since the mid-1960s \cite{BahcallWolf} that the core of a neutron star must contain it
   mesons (pions), i.e., the lightest mesons. Bose condensation of
   pions in nuclear matter is usually hampered by strong pion-nucleon repulsion. However, it was shown in
    \cite{Migdal71,Migdal77,SawyerPi,Scalapino} that
   collective excitations (pion-like quasiparticles) may arise in a
   superdense medium and condense with the loss of 
   translational invariance. Further studies revealed the possibility of
   creating different phases of the pion condensate and the
   importance of correlations between nucleons for its 
   existence. It was shown that short-range correlations and the
   formation of ordered structures in the dense matter interfere
   with pion condensation \cite{Kunihiro}.

Kaons ($K$-mesons)are the lightest strange mesons. They
   appear in the core of a neutron star as a result of the processes $e+N\to
K^-+N+\nu_e$ and $n+N\to p+K^-+N$, where $N$ is
   a nucleon whose participation ensures the momentum and
   energy conservation in the degenerate matter. The possibility
   of Bose condensation of kaons at
$\rho\gtrsim3\rho_0$ first understood in
   the 1980s 
\cite{KaplanNelson} has thereafter been studied by many authors
   (see \cite{Ramos} for a review). In a neutron star, it involves
   $K^-$-like particles, by
   analogy with pion condensation. These particles have a
   smaller mass than isolated $K$ mesons; it is this property that
   makes their Bose condensation possible. A method for the
   theoretical description of a kaon condensate taking the effects
   of strong interaction in baryonic matter into account was
   developed in \cite{Kolomeitsev}. Formation of the kaon condensate
   depends on the presence of hyperons and strongly affects the
   properties of the nucleon component of matter. Kaon
   condensation, like pion condensation, is accompanied by
   the loss of translational invariance. The condensate forms
   via first- and second-order phase transitions, depending on
   the strength of the force of attraction between kaons and
   nucleons \cite{GlendSchaff}.  Both pion and kaon condensations make the
   equation of state much softer.

\subsubsection{Quark deconfinement}

Because hadrons are made of
   quarks, the fundamental description of dense matter must
   take the quark degrees of freedom into account. Quarks
   cannot be observed in a free state when their density is low
   because they are held together (confinement) by the binding
   forces enhanced at low energies \cite{QCD}. As the density (and
   hence, the characteristic energies of the particles) grows,
   baryons fuse to form quark matter. In 1965, Ivanenko and
   Kurdgelaidze \cite{Ivanenko65} suggested that neutron stars have
   quark cores. With the advent of quantum chromodynamics,
   calculations of quark matter properties were performed in
   terms of the perturbation theory using the noninteracting
   quark model as the initial approximation
\cite{CollinsPerry,Kurkela}. However, the use of this theory is limited to energies 
 $\gg1$~GeV,
   while the chemical potential of particles in neutron stars does
   not reach such high values. More quark matter models were
   proposed, and the superfluidity of this matter associated with
   quark Cooper pairing was considered (see the references in
[\citenum{NSB1}, \S\,7.5]). These models were used to explore quark stars and
   hybrid stars, i.e., neutron stars with cores made of quark
   matter \cite{Glendenning}. For instance, the authors of
\cite{Blaschke09} predicted a series of phase
   transitions in the interior of a hybrid star with sequential
   deconfinement of quark flavors at $n_b\sim0\dec25$,
   $0\dec5$--$0\dec8$, and
$1\dec1$--$1\dec8$ fm$^{-3}$.

All published models of phase transitions in the cores of
   neutron stars have serious drawbacks. The quark and
   baryonic phases are typically treated in the framework of
   different models and cannot therefore be described self-
   consistently. Calculations from the perturbation theory are
   unrealistic at those relatively small densities at which they
   predict a phase transition. For this reason, the existence of a
   quark core in neutron stars cannot be proved theoretically.
   However, it may be hoped that this goal will be achieved
   based on the analysis of observations of compact stars.

\subsubsection{Mixed phases}

First-order phase transitions can be
   realized via a state in which one phase co-exists with another
   in the form of droplets. Such phase transitions, called noncongruent 
\cite{ilios10}, have been considered in connection with
   compact stars since the 1990s \cite{Glend92mix}.The coexistence of two
   phases in the core of a neutron star is possible thanks to the
   abandonment of the implicit assumption of electroneutrality
   of each individual phase. In the mixed phase, the electric
   charge of one component is on the average compensated by
   that of the other, and the matter structure is determined by the
   balance of surface tension at the boundaries between
   droplets, the energy density of baryonic matter, the kinetic
   energy of the constituent particles, and electrostatic energy.
   Mixed states are feasible for both meson condensation and
   baryon dissociation into quarks.

\subsubsection{Crystalline core}

The early models of neutron stars
   assumed that strong short-range neutron-neutron repulsion
   results in the formation of the solid inner core of a neutron
   star 
\cite{Cazzola},  as mentioned in review by Ginzburg
\cite{Ginzburg71}. In
   subsequent works, it was taken into account that the
   nucleon-nucleon interaction occurs by an exchange of vector
   mesons, which gives rise to the effective Yukawa potential. As
   the calculations became more exact by the late 1970s, it was
   understood that the realistic effective potentials of neutron-neutron interactions do not lead to crystallization \cite{TakemoriGuyer}.

An alternative possibility of crystallization arises from the
   tensor component of the mid-range nucleon-nucleon 
   attraction \cite{PandhaSmith}. It was shown in
\cite{TakatsukaTamagaki77}  that tensor interaction can
   lead to structures in which neutrons are located in a plane
   with oppositely oriented spins, each such plane hosting
   oppositely directed proton and neutron spins -- so-called
   alternating spin (ALS) structures. If the gain in the binding
   energy during formation of an ALS structure exceeds the loss
   of the kinetic energy of the particles, then the structure may
   become energetically favorable and a phase transition into
   this state occurs.

Moreover, given a low enough abundance of protons in
   baryonic matter, $x_p\lesssim0\dec05$, their localization may occur,
   accompanied by modulation of the neutron density
\cite{KutscheraWo89}. Under certain conditions, mixed phases may also be just
   ordered into periodic structures (see monograph \cite{Glendenning} and the
   references therein). 

In other words, there are numerous hypotheses regarding
   the structure and composition of the neutron star core,
   differing in details of microscopic interactions and 
   theoretical models for their description. We consider how they treat
   the parameters of neutron stars.
   
\subsection{Relation to observations\label{sect:coreobs}}

Manifestations of the properties of a neutron star core can be
   arbitrarily categorized into dynamic and quasistatic. The
   former are due to relatively fast processes inside the star.
   For example, a phase transition in the inner core may occur
   not only at the star birth but also as it goes through its
   evolution, e.g., during cooling (when the temperature
   decreases to below the critical one) or rotation slowdown
   (when the central pressure increases due to a reduction in
   centrifugal forces). Such a phase transition leads to a
   starquake with the release of thermal energy, a burst of
   neutrinos, excitation of crustal oscillations, and an abrupt
   change in the rotational velocity due to the altered moment of
   inertia 
\cite{HDP90}. All these effects could be possible to record and
   measure under favorable conditions. The authors of 
\cite{TakatsukaTamagaki88b}
attributed sharp jumps of pulsar rotational periods (glitches)
   to such phase transitions. Starquakes and glitches may be a
   consequence of occasional adjustment of the crust rotational
   velocity to the rotation rate of the superfluid component of
   the nucleon liquid \cite{BP71,Ginzburg71,Alpar98}.

\begin{figure}[t]
\includegraphics[width=\columnwidth]{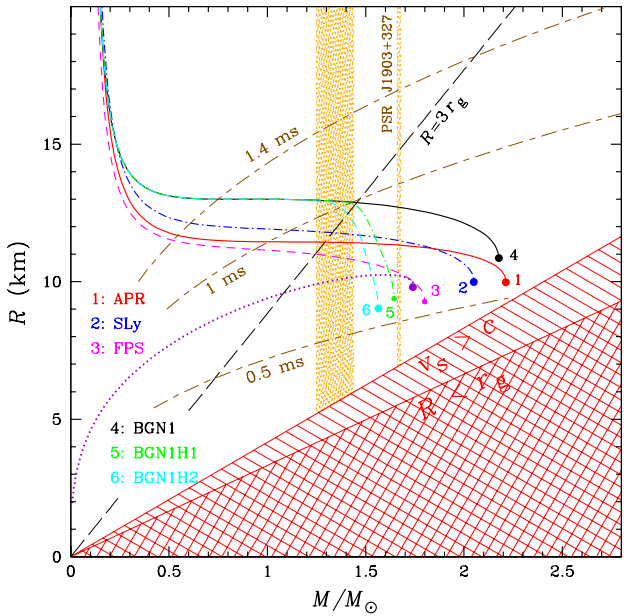}
\caption{Mass dependence of the compact star radius. Curves 1\,--\,6
correspond to the equations of state of neutron stars shown in
Figs.~\ref{fig:sly} and \ref{fig:bgn}. The dotted curve corresponds to one of the feasible equations of
   state of a quark star. The hatched triangular area is forbidden by the
   causality principle. The crosshatched triangular area lies below the event
   horizon. Three short- and long-dashed curves show additional constraints
   for rotating neutron stars: the area under the corresponding curve is
   permitted at the rotation periods indicated (1.4, 1, and 0.5 ms). The
   straight line $R=3r_g$ corresponds to the minimal stability radius of the
   circular orbit of a test particle around the neutron star of a given mass. The
   wide vertical strip encompasses the masses of binary neutron stars
   measured with an error less than $0\dec1\,M_\odot$ at confidence
   level $2\sigma$ \cite{NSB1}; the narrow
   vertical strip corresponds to the mass of the PSR J1903+0327
   millisecond
   pulsar 
\cite{Freire}. \emph{(See the Note added in proof.)}
\label{fig:RMpsr}}
\end{figure}

 Quasistatic manifestations include, inter alia, the effects
   of the core structure and the physical properties of its
   superdense matter on the theoretical radius and cooling rate
   of the star. The influence on the cooling, and hence the
   effective surface temperature, originates from the difference
   in the rates and mechanisms of neutrino losses in individual
   core models
\cite{Yak-super,Yak-neu}. The influence on the relation between the
   stellar radius $R$ and mass $M$ 
is realized via the function $P(\rho)$,
in accordance with the TOV equation. Figure~\ref{fig:RMpsr} exemplifies 
the $R(M)$ dependence for six equations of state
   of the neutron stars shown in Figs.~\ref{fig:sly} 
   and \ref{fig:bgn}, and one equation
   for a quark (strange) star (with the use of data from \cite{NSB1}).
It can be seen that quark stars must have smaller masses and
   radii than typical neutron stars. Solid dots at the ends of the
   curves correspond to the maximum mass of a stationary star
   for each equation of state. If a neutron star of a higher mass is
   discovered, the equation in question can be discarded.

Besides the equations of state, there are 
general theoretical constraints on the possible
   values of masses and radii.
   Evidently, the radius $R$ of any star must not be smaller than $r_g$; 
   otherwise, we are dealing with a black hole. Moreover, it
   can be shown \cite{NSB1} that the condition $v_s<c$, where $v_s$ is the
   speed of sound in a local reference frame and $c$ is the speed of
   light in the vacuum, imposed by the special theory of relativity
   and the causality principle, requires that the relation
$R>1\dec412\,r_g$, be satisfied, which excludes the point $(M,R)$ from
entering inside the hatched triangle in Fig.~~\ref{fig:RMpsr}.

An additional limitation is needed for the gravitation of a
   rotating star to overcome the centrifugal acceleration.
   Clearly, the radius must not be too large. In Fig.~\ref{fig:RMpsr}, the
   largest values of the radius in the dependence on the mass
$M$ at a given rotation period $P$ are shown by alternating short
and long dashes for $P=1$ ms, $1\dec4$ ms, and
$0\dec5$ ms.  We see that the period
$P=1\dec4$ ms (the shortest of the periods observed to date)
 does not place any
   serious constrains on 
$R$. On the other hand, the pulsar period 0.5 ms is
   incompatible with any of the known theoretical equations of
   state of dense matter (the detection of such a period in the
   radiation from the 1987A supernova remnant was reported in
   1989 \cite{Kristian89},
but it later proved to be a technical error \cite{Kristian91}).

The wide vertical strip in Fig.~\ref{fig:RMpsr}
depicts the range of exactly
   measured masses of neutron stars in binary systems made up
   of a pulsar and another neutron star The narrow vertical strip
   corresponds to the estimated mass of PSR J1903+327
   mentioned in Sect.~\ref{sect:GR}. Confirmation of this estimate
   would make the choice between theoretical models much
   more definitive. For example, it follows from the figure that
   the existence of a star of such mass implies the absence of
   hyperons in the model \cite{BGN}.

Were $R$ and $M$ known exactly for a certain compact star,
   it would probably permit choosing one of the equations of
   state as the most realistic one. Unfortunately, the current
   accuracy of measurement of neutron star radii leaves much to
   be desired. Determination of stellar masses and radii requires
   a reliable theoretical description of the envelopes that
   influence the star surface temperature and the formation of
   the emitted radiation spectrum. We return to this issue in
   Sect.~\ref{sect:RM}.

\section{Envelopes\label{sect:envel}}

\begin{figure*}
\begin{center}
\includegraphics[width=.7\textwidth]{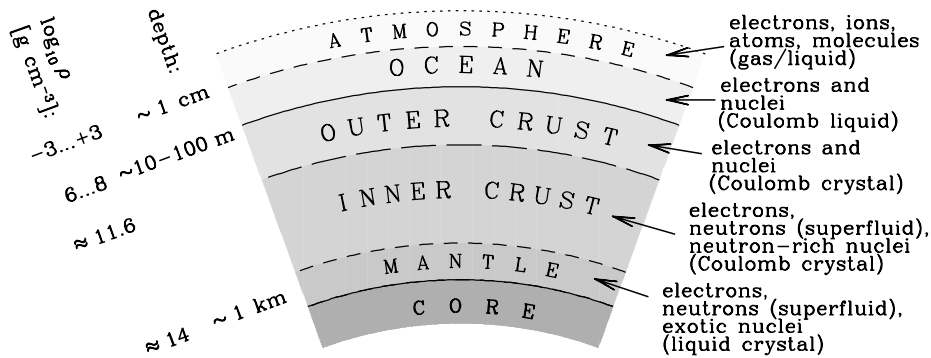}
\caption{Schematic cross section of neutron star envelopes. From bottom
   up: core, mantle, inner crust, outer crust, ocean, atmosphere. Right-hand
   side: composition of these layers; left-hand side: characteristic values of
   density logarithm and depth from the surface.
}
\label{fig:ns-str}
\end{center}
\end{figure*}

Envelopes of a neutron star are divisible into the solid crust in
   which atomic nuclei are arranged into a crystal and the liquid
   ocean composed of the Coulomb fluid. The crust is 
   subdivided into the inner and outer parts. In the former, the
   nuclei are embedded in a sea of free neutrons and electrons,
   while the latter contains no free neutrons. A neutron star can
   have a gaseous plasma atmosphere at the surface, while
    the stellar core may be surrounded by a liquid crystal
   mantle topped by the crust.

\subsection{Inner crust\label{sect:inner}}

\emph{The inner crust} s normally $\sim1$\,--\,$2$ km thick. Its density
   increases from 
$\rho_\mathrm{drip}\approx(4$\,--\,$6)\times10^{11}$ \gcc, , at which 
   neutrons begin to ``drip'' from the nuclei, to
$\sim0\dec5\rho_0$, when the atomic
 nuclei fuse into a homogeneous mass. The nuclear chemical
   equilibrium with respect to beta-capture and beta-decay
   reactions in the inner crust accounts for the matter 
   composition that cannot be reproduced under laboratory conditions
   (neutron-rich heavy nuclei embedded in a fluid composed of
   neutrons and electrons). The physics of such matter is fairly
   well described in 
\cite{PethickRavenhall}. The neutrons in a large portion of the
   inner crust are superfluid; according to theoretical estimates,
   the critical superfluidity temperature varies with density and
   reaches billions of degrees or an order of magnitude higher
   than the typical kinetic temperature of matter in the inner
   crust of a neutron star.

The pressure in the inner crust of a neutron star is largely
   created by degenerate neutrons. However, superfluidity may
   decrease their heat capacity and is therefore responsible for
   the decisive contribution of atomic nuclei to the thermal
   capacity of the inner crust. The nuclei make up a crystal
   lattice, formed essentially by Coulomb interaction forces
   (Coulomb or Wigner crystal). An adequate description of
   their contribution is possible by considering collective
   vibrational excitations (phonon gas). Because the electrons
   are relativistic and highly degenerate particles, their 
   contribution to the heat capacity of the inner crust is insignificant
   unless the temperature is too low. However, it may 
   appreciably increase when the temperature of the Coulomb crystal
   decreases much below the Debye temperature, imposing a
   ``freeze-out'' on phonon excitations [\citenum{NSB1}, \S\,2.4.6].

The electric conductivity in the inner crust is due largely to
   electrons, whose scattering plays an important role. The
   scattering on phonons of a crystal lattice prevails at relatively
   high temperatures, and that on lattice defects or admixtures is
   responsible for the residual resistance at low temperatures.
   Ions (atomic nuclei) incorporated into the crystal lattice make
   no appreciable contribution to conductance. At the same
   time, thermal conductivity is due to phonons and neutrons,
   besides electrons whose scattering is governed by the same
   mechanisms that operate in the case of electric conductivity
   supplemented by electron-electron collisions. Phonons may
   become the main heat transfer agents in the presence of lattice
   defects and admixtures hampering the participation of
   electrons in this process \cite{ChugunovHaensel}.
Neutrons, especially 
   superfluid ones, may also serve as heat carriers in the inner
   crust \cite{Aguilera-neutrons}.

\subsection{Mantle\label{sect:mantle}}

The core of a neutron star may be separated from the bottom of its
   inner crust by a layer that contains exotic atomic nuclei and is
   called the mantle 
\cite{PethickPotekhin}. In the liquid-drop
model, the spherical shape of the atomic nucleus is 
   energetically advantageous at a low density,
   since it minimizes the surface energy. However, the
   contribution from the Coulomb energy at a higher 
   density may change the situation. The mantle consists of a few
   layers containing such phases of matter in which atomic
   nuclei are shaped not like spheres but rather likes cylinders
   (the so-called ``spaghetti'' phase), plane-parallel plates
   (``lasagna'' phase), or ``inverse'' phases composed of nuclear
   matter with entrapped neutron cylinders (``tubular'' phase)
   and balls (``Swiss cheese'' phase) \cite{PethickRavenhall}. 
   Such structures for collapsing cores of supernovae were 
   first conjectured in 
\cite{RPW83}, and for neutron stars in \cite{Lorenz93}. 
Whereas the spherical nuclei constitute a 3D crystalline lattice,
the mantle has the properties of a liquid crystal
\cite{PethickPotekhin}.
Direct Urca processes of
   neutrino emission can be allowed in the mantle \cite{Gusakov_ea04},
   while they are
   unfeasible in other stellar envelopes; their high intensity
can enhance the cooling of the neutron star.

Not all current equations of state of nuclear matter predict
the mantle; some of them treat such a state as energetically
   unfavorable. The mantle hypothesis appears in the FPS
   model, but not in the most modern Sly model.

\subsection{Outer crust and its melting}

The outer shells of a neutron star are hundreds of meters thick
   and consist of an electron-ion plasma that is completely
   ionized, that is, consists of ions in the form of atomic nuclei
   and strongly degenerate free electrons (probably except a
   several-meter-thick outer layer with the density below
   $10^6$ \gcc). Then the total pressure is determined by the pressure
   of degenerate electrons. The electrons become relativistic
   (with the Fermi momentum $p_\mathrm{F}$ comparable to $m_ec$, where
   $m_e$
   is the electron mass) 
   at $\rho\gtrsim10^6$ \gcc\ and ultrarelativistic
 ($p_\mathrm{F}\gg m_e c$) at $\rho\gg10^6$ \gcc. At such densities, ions give rise
   either to a Coulomb liquid (whose properties mostly depend
   on Coulomb interactions between ions) or to a Coulomb
   crystal.

The electron Fermi energy in deep-lying layers of the outer
   envelopes increases so as to enrich nuclei with neutrons by
   virtue of beta-captures. Finally, the inner-outer crust 
   interface forms at $\rho=\rho_\mathrm{drip}$, where free neutrons appear.

     The external boundary of the outer crust normally
   coincides with the crystallization point of the Coulomb
   liquid making up the neutron star ocean. The position of
   this point is given by the density dependence of the Coulomb
   crystal melting temperature. In the so-called one-component
   Coulomb plasma model disregarding electron-ion 
   interactions and treating ions as classical point particles, formation
   of the Coulomb crystal is defined by the equality $\Gamma=175$ or
   (in a more realistic approach) $\Gamma\sim100$\,--\,200
\cite{PC00}. Here, $\Gamma=(Ze)^2/(a\kB T)$is the Coulomb coupling parameter
   characterizing the relation between the potential Coulomb
   energy of the ions and their kinetic energy, $a=(4\pi
n_i/3)^{-1/3}$ is
   the ion-sphere radius, $n_i$ is the ion volume density, and
$\kB$ is the Boltzmann constant. The melting point for a typical
   neutron star envelope lies at $\rho_\mathrm{m}\sim10^6-10^9$ \gcc, 
   depending on its thermal structure (i.e., temperature variations with
   depth related to the star age and past history). However, a
   cold enough neutron star may lose both the atmosphere and
   the ocean in a superstrong magnetic field; in such a case, the
   external boundary of the crust coincides with the stellar
   surface (see \cite{Lai01} for discussion and references). 

\subsection{Ocean}

The bottom of the neutron star ocean is located at the melting
   point with the density $\rho_\mathrm{m}$, while its surface is arbitrary because
   of the lack of a clear-cut ocean-atmosphere interface on a
   typical star. An exception, as in the case of the solid crust, is
   neutron stars having a rather strong magnetic field, which
   may be responsible for the absence of an optically thick
   atmosphere and its substitution by a liquid boundary. Most
   of the ocean consists of atomic nuclei surrounded by
   degenerate electrons. Therefore, in the general case, we
   speak of ions surrounded by electrons, with the 
   understanding that ions mean both completely and partially ionized
   atoms.
   
The ocean matter is the Coulomb liquid, most of which is
   strongly coupled, i.e., 
$\Gamma\gg1$. 
One of the main problems in
   theoretical studies of such matter is adequate consideration of
   the influence of microscopic correlations between ion 
   positions on the macroscopic physical characteristics of the
   matter being investigated, such as equations of state \cite{PC00}
   and kinetic coefficients \cite{BKPY}.

\subsection{Atmosphere}

The stellar atmosphere is a layer of plasma in which the
   thermal electromagnetic radiation spectrum is formed. The
   spectrum contains valuable information about the effective
   surface temperature, gravitational acceleration, chemical
   composition, magnetic field strength and geometry, and
   mass and radius of the star. The geometric thickness of the
   atmosphere varies from a few millimeters in relatively cold
   neutron stars (effective surface temperature
$T_\mathrm{eff}\sim10^{5\dec5}$~K) to
   tens of centimeters in rather hot ones
($T_\mathrm{eff}\sim10^{6\dec5}$~K). In most
   cases, the density of the atmosphere gradually (without a
   jump) increases with depth; however, as mentioned above,
   stars with a very low effective temperature or superstrong
   magnetic field have either a solid or a liquid condensed
   surface.

The deepest layers of the atmosphere (its ``bottom'' being defined
   as a layer with the optical thickness close to unity for the
   majority of outgoing rays) may have the density $\rho$ from $\sim10^{-4}$
to $\sim10^6$ \gcc,  depending on the magnetic field $B$,
temperature $T$, gravity $g$, and the
   chemical composition of the surface. The presence in the
   atmosphere of atoms, molecules, and ions having bound
   states substantially alters absorption coefficients of 
   electromagnetic radiation and thereby the observed spectrum.

     Although the neutron star atmosphere has been 
     investigated by many researchers for several decades, these studies
   (especially concerning strong magnetic fields and incomplete
   ionization) are far from being completed. For magnetic fields
$B\sim10^{12}$\,--\,$10^{14}$~G, this problem is practically solved only for
   hydrogen atmospheres with
$\Teff\gtrsim10^{5\dec5}$~K \cite{KK,HoPC07}. The
   bound on $\Teff$ from below is related to the requirement of
   smallness of the contribution from molecules compared to
   that from atoms, the quantum mechanical properties of
   molecules in a strong magnetic field being poorly known.
   For
$B\sim10^{12}$\,--\,$10^{13}$~G and $10^{5\dec5}$~K
$\lesssim\Teff\lesssim 10^{6}$~K, there are
   models of partly ionized atmospheres composed of carbon,
   oxygen, and nitrogen \cite{MoriHo}. Here, the restriction on $\Teff$ from
   above arises from the rough interpretation of ion motion
   effects across the magnetic field that holds at low thermal
   velocities (see Sect.~\ref{sect:magnatom}).

\section{Magnetic fields\label{sect:mag}}

\subsection{Magnetic field strength and evolution}

As discussed in the Introduction, the majority of currently
   known neutron stars have magnetic fields unattainable in
   terrestrial laboratories, with typical values
$B\sim10^8$\,--\,$10^{15}$~G
   at the surface, depending on the star type. The field strength
   inside a star can be even higher. For example, certain
   researchers propose explaining the energetics of AXP and
   SGR in terms of core magnetic fields as high as
    $B\sim10^{16}$\,--\,$10^{17}$~G at the birth of the neutron star (see 
    \cite{DallOsso_ss} and the
   references therein). The theoretical upper bound obtained
   numerically in 
\cite{Bocquet} 
is consistent with the estimate from the
   virial theorem \cite{ChandraFermi,LaiShapiro91}:
 $\max(B)\sim10^{18}$~G.

A number of theoretical models of field generation have
   been proposed suggesting the participation of differential
   rotation, convection, magneto-rotational instability, and
   thermomagnetic effects either associated with supernova
   explosion and collapse or occurring in young neutron stars
   (see 
\cite{Reisenegger}). . Specifically, the
``$\alpha$--$\Omega$-dynamo model'' \cite{Bisno92dynamo,TD93}
assumes that the core of a neutron star born with a sufficiently
   short (millisecond) rotation period acquires a toroidal
   magnetic field up to 
$B\sim10^{16}$~G due to differential rotation,
   while the pulsar magnetic field is generated by means of
   convection at the initial rotation periods
$\gtrsim30$~ms. However,
 none of the proposed models is able to account for the totality
   of currently available neutron star data.

Electric currents maintaining the stellar magnetic field
   with the involvement of differential rotation circulate either in
   the inner crust or in the core of a neutron star, i.e., where
   electric conductivity is high enough to prevent field decay for
   a time comparable with the age of known pulsars. It was
   shown as early as 1969 \cite{BaymPP} that the characteristic time of
Ohmic decay of the core magnetic field may exceed the age of
   the Universe. For a magnetic field originating in the core of a
   neutron star, proton superconductivity stipulates its
   existence in the form of quantized magnetic tubes 
   (Abrikosov vortices, or fluxoids) having a microscopic transverse dimension.

The magnetic field of a neutron star changes in the course
   of its evolution, depending on many factors and inter-related
   physical processes (see, e.g., 
\cite{Cumming_ea04} and the references therein).
   Specifically, the field undergoes ohmic decay and a change in
   configuration under the effect of the Hall drift; also, magnetic
   force lines reconnect during starquakes. Thermoelectric
   effects, as well as the dependence of components of the
   thermal and electrical conductivity tensors, and
   plasma thermoelectric coefficients on temperature and the
   magnetic field, are responsible for the interrelation between
   magnetic and thermal evolution \cite{UrpinKonenkov08,Pons_MG}.
Accretion can also strongly affect
the near-surface magnetic field \cite{Bisno06,Cumming_ea04}. 

The evolution of a magnetic field generated by Abrikosov
   vortices is to a large extent dependent on their interaction
   with other core components, such as Feynman--Onsager
   vortices in the neutron superfluid \cite{BaymPP,GinzburgKirzhnits},
and conditions at
   the core boundary, i.e., interactions of these vortices with
   crustal matter \cite{Alpar98}.

\subsection{Landau quantization}

The motion of free electrons perpendicular to the field is quantized
   into Landau levels \cite{Landau30}. Their characteristic transverse scale
   is the magnetic length
$\am=(\hbar c/eB)^{1/2}$, and the inter-level
   distance in a nonrelativistic theory is the cyclotron energy
$\hbar\omc=11\dec577\,B_{12}$ keV, where
$\omc = eB/(m_e c)$ is the electron
   cyclotron frequency (the notation
$B_{12}=B/(10^{12}\textrm{~G})$is 
   introduced here). The dimensionless parameters characterizing a
   magnetic field in relativistic units $b$ and atomic units $\gamma$
   are 
\bea 
 b&=&
{\hbar\omc}/({m_e c^2}) = 
{B_{12}}/{44\dec14} \,,
\\
\gamma&=&\bigg(\frac{\aB}{\am}\bigg)^{\!2}
\!\! = \frac{\hbar\omc}{2\,\mathrm{Ry}} = \frac{\hbar^3
B}{m_e^2\,c\,e^3} = 425\dec44\,B_{12} \,, 
\eea
where $\aB$ is the Bohr radius.
We call a magnetic field strong if $\gamma\gg1$
and superstrong if $b\gtrsim1$. In the relativistic theory, the
   energies of Landau levels are $E_N=m_e c^2
\,(\sqrt{1+2bN}-1)$ ($N=0,1,2,\ldots$). 

In a superstrong field, specific effects of quantum 
     electrodynamics, such as electron-positron vacuum polarization in
   an electromagnetic wave field, become significant. As a result,
   the vacuum acquires the properties of a birefringent medium,
   which, at $b\gtrsim1$, markedly affects the radiation spectrum
   formed in the atmosphere of a neutron star
\cite{PavlovGnedin,HoLai02}.

For ions with charge $Ze$ 
and mass $\mion=Am_u$, where $m_u=1\dec66\times10^{-24}$ g
is the unified atomic mass unit,
the cyclotron frequency equals $\omci = |Ze|B/(\mion
c)$, the cyclotron energy
$\hbar\omci=6\dec35\,(Z/A)\,B_{12}$ eV, 
and the parameter that characterizes the role of relativity effects 
$b_i=\hbar\omci/(\mion
c^2)=0\dec68\times10^{-8}\,(Z/A^2)\,B_{12}$. The smallness of $b_i$ 
allows one to ignore the relativistic effects for ions in the atmosphere of a neutron
   star.

The motion of electrons along a circular orbit in a
   magnetic field in the classical theory leads to cyclotron
   radiation at the frequency
$\omc$ and the formation of a
   cyclotron line in the magnetic atmosphere. In quantum
   theory, the cyclotron line corresponds to transitions between
   the adjacent Landau levels. Transitions between distant
   Landau levels result in the formation of cyclotron harmonics
   with energies $E_N$. The discovery of the first cyclotron line
with the energy 58 keV in the spectrum of the X-ray pulsar in the
binary Hercules X-1 in 1978 
\cite{Truemper78}
gave a stunning argument in favor of
the idea of neutron star magnetic fields. A number of similar
systems with synchrotron lines are known  presently. The spectra
of several X-ray pulsars in binary systems exhibit cyclotron
harmonics with $N\geqslant1$ (see \cite{RodesRoca,Enoto}); the
observation of up to four harmonics has been reported
\cite{Santangelo}.

The spectra of isolated neutron stars may likewise contain
   cyclotron lines. Electron cyclotron lines can be seen in the
   thermal spectral range from 0.1 to 1 keV at $B\sim10^{10}$\,--\,$10^{11}$ G
   and ion cyclotron lines at $B\sim10^{13}$\,--\,$10^{14}$
G. There is a hypothesis that
   absorption lines in the spectrum of CCO 
1E 1207.4--5209 can be explained by the cyclotron mechanism
\cite{Bignami,SuPaW}. 
   It should be noted that ion cyclotron harmonics, unlike electron
   ones, are too weak to be observed \cite{P10}.

The effect of Landau quantization on plasma properties is
   significant when the cyclotron energy is not too small
   compared with the thermal ($\kB T$) and Fermi ($\EF$) energies. 
   If
$\hbar \omc$ is much higher than these two energies, most electrons in
   thermodynamic equilibrium are at the ground Landau level.
   In this case, the field is called \emph{strongly quantizing}. 
   If, at contrast, $\kB T$ or $\EF$is much greater than the energy difference between the
   adjacent Landau levels, the field is nonquantizing. The
   smallness condition on the thermal energy compared with the
   cyclotron one can be written as $\hbar\omc/(\kB
T) \approx 134\,B_{12}/T_6 \gg1$ for the electrons and $\hbar\omci/\kB T
\approx 0\dec0737\,(A/Z) B_{12}/T_6\gg1$ for the ions. The second condition
(higher $\hbar\omc$ than $\EF$) imposes a
   bound on density. Degenerate electrons occur at the ground
   Landau level if their number density
$n_e$ is smaller than $n_B\equiv(\pi^2\sqrt2\,\am^3)^{-1}$.
Therefore, the field will be strongly quantizing at $\rho<\rho_B$,
where
\beq 
\rho_B = \mion n_B/Z \approx
7\times10^3 \,(A/Z) \,B_{12}^{3/2}\text{ \gcc}. 
\label{rho_B}
\eeq
 At $\rho>\rho_B$ the field is weakly quantizing, and at
$\rho\gg\rho_B$ it can be treated as nonquantizing. 
Estimates, analogous to (\ref{rho_B}), are feasible also for other
fermions
[\citenum{NSB1}, \S\,5.17].

We note that a nonquantizing magnetic field has no effect
   on the equation of state (Bohr -- van Leeuwen theorem). It
   follows from Fig.~\ref{fig:ns-str} with account of (\ref{rho_B}) that
   magnetic fields inherent in neutron stars are strongly
   quantizing in the atmosphere and can be quantizing in both
   the ocean and the outer crust; however, even at limit values
$B\sim10^{18}$~G, the magnetic fields do not affect the stellar core
   equation of state. These conclusions are based on simple
   estimates, but they are confirmed by calculations of the
   nuclear matter equation of state in superstrong magnetic
   fields \cite{Broderick}.

\subsection{Atoms and ions in magnetic atmospheres}
\label{sect:magnatom}

The atmosphere of a neutron star contains atoms, molecules,
   and atomic and molecular ions having bound states. Strong
   magnetic fields markedly affect their quantum mechanical
   properties (see reviews 
\cite{JHY83,Ruder,Lai01}). It was suggested soon
   after the discovery of pulsars \cite{Ruderman71},
that at equal temperatures,
   there should be more atoms in the neutron star atmosphere at $\gamma\gg1$
than at $\gamma\lesssim1$, because in a strong magnetic field, the
   binding energies of their ground state and a certain class of
   excited states (so-called \emph{tightly bound states}) markedly
   increase and the quantum mechanical size decreases.
For instance, the ground-state energy of an H atom at
 $B\sim10^{11}$\,--\,$10^{14}$ G can be roughly estimated as $E\sim 200\,(\ln
B_{12})^2$~{eV}.
 In all the states at $\gamma\gg1$, the electron cloud acquires the form of an
   extended ellipsoid of rotation with the characteristic small
   semiaxis
$\sim\am=\aB/\gamma$ and large semiaxis $l\gg\am$
($l\sim\aB/\ln\gamma\gg\am$ for the tightly-bound states). 
Accurate fitting formulae for energies and other characteristics
of a hydrogen atom in magnetic fields are given in 
\cite{P98}.

The properties of molecules and even the very existence of
   some of their types in strong magnetic fields are poorly
   known, although they have been discussed for almost 40
   years. Those diatomic molecules are fairly well studied whose
   axis coincides with the direction of the magnetic field. For
   obvious reasons, the H$_2$molecule has been thoroughly
   investigated. The approximate formulas for its binding
   energy at $\gamma\gtrsim10^3$
   increasing at the same rate $\propto(\ln\gamma)^2$
   as the
   binding energy of H atoms are presented in 
\cite{Lai01}. 
Interestingly, however, numerical calculations in \cite{Detmer}, show that this
   molecule is unstable in a moderate magnetic field (in the range
    $0\dec18 < \gamma < 12\dec3$).

Also, the  H$_2$$^+$ion has been studied fairly well
   (see, e.g.,
\cite{KS96});
HeH$^{++}$, H$_3$$^{++}$, and other exotic one-electron molecular
   ions becoming stable in strong magnetic fields were 
  also considered
\cite{Turbiner07}. 

A strong magnetic field can stabilize polymer molecular
   chains aligned along magnetic field lines. These chains can then
   attract one another via dipole-dipole interactions and
   make up a condensed medium. Such a possibility was first
   conjectured by Ruderman in 1971
\cite{Ruderman71}. Investigations in the
   1980s --
2000s showed that in the fields $B\sim10^{12}$--$10^{13}$~G
these chains are formed not of any chemical elements,
but only of the atoms from  H to C, and undergo polymerization into a condensed
   phase either in a superstrong field or at a relatively low
   temperature, with the sublimation energy of such condensate
   being much smaller than predicted by Ruderman (see \cite{MedinLai06b}  and the references therein).

The overwhelming majority of researchers of atoms and
   molecules in strong magnetic fields have considered them to
   be at rest. Moreover, in studies of electron shells, the atomic
   nuclei were almost universally assumed to be infinitely
   massive (fixed in space). Such an approximation is a gross
   simplification for magnetic atmospheres. Astrophysical
   simulations must take the finite temperature and therefore
   thermal motion into account. Atomic motion across magnetic
   field lines breaks the axial symmetry of a quantum 
   mechanical system. At
$\gamma\gg1$ , specific effects associated with the
   collective motion of a system of charged particles become
   significant. Specifically, the \emph{decentered states} can be 
   populated in which electrons are mainly located in a ``magnetic
   well''
   far from the Coulomb center. These exotic states were
   predicted for hydrogen atoms in \cite{Burkova}.
   In the same paper, and later in
\cite{IPM84}, the first studies of the energy spectrum of these states
were performed. 

 Even at low temperatures, when the thermal motion of
   atoms can be neglected in the first approximation, the finite
   mass of the atomic nucleus should be taken into 
   consideration, even in a sufficiently strong field; the nucleus undergoes
   oscillations in a magnetic field due to Landau quantization
   even if the generalized momentum 
\cite{JHY83} describing the motion
   of the center of masses across the field is zero. Different
   quantum numbers of an atom correspond to different
   vibrational energies that are multiples of the cyclotron
   energy of the atomic nucleus. In a superstrong field, this
   energy becomes comparable to the electron shell energies and
   cannot therefore be disregarded.

A comprehensive calculation of hydrogen atom energy
   spectra taking account of motion effects across the strong
   magnetic field was carried out in
\cite{VDB92,P94}, and the
   calculation of the probability of different types of radiative
   transitions and absorption coefficients in neutron star 
   atmospheres in a series of studies was reported in 
\cite{PC03}. Based on
   these data, a model of the hydrogen atmosphere of a neutron
   star with a strong magnetic field \cite{KK} was elaborated. The
   database for astrophysical calculations was created using this
   model in \cite{HoPC07}.

The quantum mechanical effects of He+ ion motion were
   considered in \cite{BPV97,PB05}. This case is essentially different from
   that of a neutral atom in that the values of the ion generalized
   momentum are quantized \cite{JHY83}.
For many-electron atoms,
   molecules, and ions, the effects of motion across the magnetic
   field remain unexplored. The perturbation theory applicable
   to the case of small generalized momenta
\cite{VB88,PM93} may prove
   sufficient to simulate relatively cold atmospheres of neutron
   stars 
\cite{MoriHo}. 

\subsection{Electron heat and charge transport coefficients}

The magnetic field affects the kinetic properties of the plasma
   in a variety of ways (see, e.g.,\cite{YaK94}, for a review).
Any magnetic field makes the
   transfer of charged particles (in our case, electrons) 
   anisotropic. It hampers their motion and thereby heat and charge
   transfer by electrons in the direction perpendicular to the
   field, thus generating Hall currents. These effects are essential
   when the cyclotron frequency $\omc$ is much higher than the
   effective collision frequency, while the latter remains 
   unaltered in a nonquantizing field.

\begin{figure}[t]
\begin{center}
\includegraphics[width=\columnwidth]{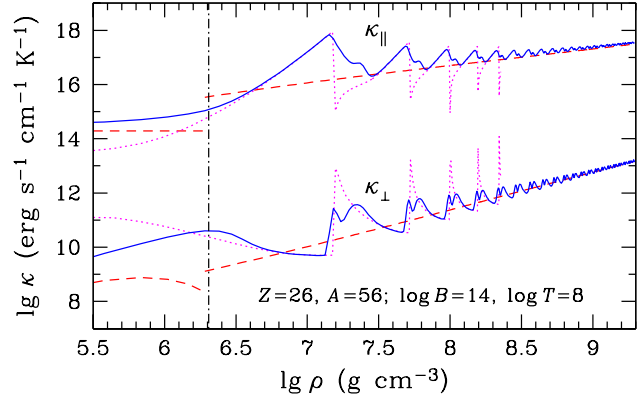}
\end{center}
\caption{Longitudinal ($\|$) and transverse ($\perp$) thermal
conductivities in the iron outer
   envelope of a neutron star at
$T=10^8$~K and $B=10^{14}$~G \cite{VP01}. Solid
   curves are calculations according to \cite{P99}, dashed lines are the classical
   model, the dotted line is the results neglecting thermal averaging.
   Electrons are degenerate to the right of the vertical dashed-dotted line,
   whose position corresponds to the equality $\TF=T$.
\label{fig:elounda}}
\end{figure}

A quantizing magnetic field exerts a more pronounced
   influence on the transfer process. In a weakly quantizing field
   (in the presence of degeneracy), kinetic coefficients oscillate
   with variations of matter density about the values they would
   have in the absence of quantization. In a strongly quantizing
   magnetic field, the values of the kinetic coefficients are
   substantially different from classical ones.

The effects of quantizing magnetic fields on electron
   transfer in plasma have been studied by different authors for
   many decades (see the references in \cite{YaK94}). The formulas for
   electron kinetic coefficients of a completely ionized plasma
   convenient to use in astrophysics at arbitrary $\rho$, $B$, and $T$
   were
   derived in \cite{P99}. They were used to calculate thermal
   evolution of neutron stars as described in the next section.

Figure~\ref{fig:elounda}
exemplifies the dependences of heat conductivity
   coefficients along and across a magnetic field at the plasma
   characteristics inherent in the blanketing shell of a
   neutron star with field
$B=10^{14}$~G; the accurately computed
   characteristics (solid curves) are compared with the simplified
   models used in astrophysics previously. The dashed lines
   represent models disregarding Landau quantization under
   the assumption of a strong electron degeneracy (to the right of
   the vertical dashed-dotted line) or nondegeneracy (to the left
   of the vertical line). The dotted line is the assumption under
   which thermal scattering of electron energies near the Fermi
   energy is neglected (see \cite{VP01} for the details and references).

\section{Cooling and thermal radiation\label{sect:cool}}

\subsection{Cooling stages}

About 20 s after its birth, a neutron star becomes transparent
   to neutrino emission
\cite{ImshennikNadyozhin}, carrying away the energy to outer
   space and cooling the star. Soon after that, the temperature
   distribution in the stellar core characterized by high heat
   conductivity reaches equilibrium, preserved thereafter
   throughout the star lifetime (probably except for short
   periods after catastrophic phase transitions in the core
   postulated by certain hypothetical models). In line with GR,
   the equilibrium temperature increases toward the center of
   the star in proportion to $\mathrm{e}^{-\Phi}$, where $\Phi$ is the metric
   function defined by the star hydrostatic model and related to
   the time-component of the metric tensor 
$g_{00}=\mathrm{e}^{2\Phi}$, which decreases from the surface to the
center \cite{Thorne77}.

The stellar crust is for some time hotter than the core. The
   cooling wave reaches the surface within 10--100 years;
   thereafter, the star cools down in the quasistationary regime
   where the temperature distribution in the heat-insulating blanket
   at each time instant unambiguously depends on the core
   temperature. We note that all currently observed neutron
   stars are at least several centuries old. This means that they
   are in the state of quasistationary cooling in the absence of
   fast energy release in the envelopes. The quasistationarity
   may be disturbed by the explosive thermonuclear burning of
   accreted matter \cite{StrohmayerBildsten} or the liberation of energy in the crust
   during starquakes
\cite{Blaes-ea89,HDP90,Alpar98,Franco-ea}.

Cooling in the quasistationary regime goes through the
   following stages \cite{YakovlevPethick}.
\begin{enumerate}
\item
The \emph{neutrino cooling} stage lasts $\sim10^5$ years. During
   this time, the core cools largely via neutrino emission in
   various physical reactions \cite{Yak-neu}, the main ones being direct (if
   present) and modified Urca processes (depending on the
   particles involved), as well as neutrino bremsstrahlung
   radiation.
\item
The \emph{photon cooling} stage is the final one. It begins at
the stellar age $t\gtrsim10^5$ years when the lowered core temperature makes
   neutrino emission (strongly temperature-dependent) weaker
   than in cooling via heat transfer through the envelope and
   conversion into surface electromagnetic radiation.
\end{enumerate}

The cooling curve of a neutron star depends on its mass $M$;
the model of superdense matter in the core determining the
   equation of state (hence, the radius $R$)  and composition of the
   core (hence, the intensity of neutrino emission at a given
   mass); and the envelope properties: (a) thermal conductivity
   determining $L_\gamma$ at a given core temperature, (b) neutrino
   luminosity in the crust, (c) sources of heating and their
   intensity. Characteristic thermal conductivity and neutrino
   luminosity of the envelopes at each time instant $t$ (i.e., at the
   model-specific temperature $T$ distribution in the envelopes),
   in turn, depend on the stellar mass $M$, the radius $R$, and the
   the magnetic field (both magnetic induction
$B$ and the
   configuration of magnetic force lines may be essential).

Comparing the observed  $L_\gamma$ and $t$ for neutron stars with
   the cooling curves allows estimating $M$ and $R$
and placing
   bounds on the theoretical models of superdense matter. This
   method for parameter evaluation is largely applicable to
   \emph{isolated neutron stars}. By contrast, most neutron stars in
   binary systems have an additional source of energy (accretion) and an additional source of X-ray radiation (accretion
   disk), often much more powerful than
$L_\gamma$.

\subsection{Thermal structure}

The complete set of equations describing the mechanical and
   thermal structure and evolution of a spherically symmetric
   star at hydrostatic equilibrium in the framework of GR was
   obtained by Thorne \cite{Thorne77}.
These equations are easy to
   transform to the form holding for the stellar envelope with
   radial heat transfer, a smooth temperature distribution over
   the surface, and a force-free magnetic field. Under the
   assumption that heat transfer and neutrino emission
   are quasistationary, these equations reduce to a system of
   ordinary differential equations for the metric function
$\Phi$,    local density of the radial heat flow $F_r$,
temperature $T$ and gravitational mass $m$ comprised inside
a sphere of radius $r$,
as functions of pressure $P$ (e.g.,  \cite{PCY07}).
GR correction factors in this system depend on the mass fraction
 $(M-m)/M$outside the equipotential surface being 
   considered and on the
$P/(\rho c^2)$ ratio entering the TOV equation. At
   the outer--inner crust interface, $(M-m)/M\sim10^{-5}$ and
$P/(\rho c^2)\sim10^{-2}$, therefore, the GR correction factors
in the outer
   shells are almost constant. Bearing this constancy in mind and
   disregarding the geometric thickness of the heat-insulating
   layer compared with $R$,  in the absence of heat sources and
   sinks in the blanketing envelope of a neutron star, we
   obtain that the radial heat flow $F_r$
is constant and equal to $\sigma_\mathrm{SB}T_s^4$, where
$\sigma_\mathrm{SB}$ is the Stefan--Boltzmann constant. Here and hereafter, we discriminate
   between the local surface temperature $\Ts$
   and integral effective temperature
$\Teff$, because $\Ts$ may vary over the
   surface. Under the above conditions, calculating the thermal
   structure amounts to solving a simplified equation
   \cite{GPE}, 
   which can be written in the same form as the nonrelativistic equation
$
   \kappa\,\dd T/\dd P =
    F_r / (\rho g).
$
Such an approximation is used in the
   majority of neutron star cooling research \cite{YakovlevPethick}. 
However, since magnetars
   have stronger magnetic fields and surface luminosities than
 ordinary neutron stars and have, in addition, internal sources
   of energy, one has to solve for them the complete set of equations,
   taking neutrino emission rate per unit volume $Q_\nu$ 
and heat sources $Q_{h}$ into account, instead of the simplified heat
   transfer equation \cite{PCY07,Kaminker_ea09}.

The effective radial thermal conductivity at a local surface
   area in a magnetic field is
$
   \kappa=\kappa_\|\cos^2\theta_B + \kappa_\perp\sin^2\theta_B,
$
where $\theta_B$ is the angle between magnetic force lines and the normal to
   the surface, and $\kappa_\|$ and $\kappa_\perp$ are components of the heat
   conductivity tensor responsible for the transfer along and
   across the force lines. In the heat-insulating envelope of a
   neutron star each of the components $\kappa_\|$ and $\kappa_\perp$
contains
   radiative $\kappa_{r}$ and electron  $\kappa_{e}$ constituents. Photon heat 
   conduction prevails 
($\kappa_{r}>\kappa_{e}$) in the outermost (typically 
   nondegenerate) layers, while electron heat conduction plays the main
   role in deeper, moderately or strongly degenerate layers. The
   total thermal flux along a given radius $r$ (the local luminosity
   related to thermal but not neutrino losses) is defined by the
   flux density integral over the sphere of this radius,
$
   L_r = \int\sin\theta\dd\theta\,\dd\varphi\,
    r^2 F_r(\theta,\varphi),
$
where $(\theta,\varphi)$ are respectively
   the polar and azimuthal angles.

The use of equations holding for a spherically symmetric
   body at each point of the surface assumes that the mean radial
   temperature gradient is much greater than the lateral one. The
   estimates made in \cite{PCY07}indicate that this condition is fulfilled
   with a good accuracy at the largest part of the star surface,
   and corrections for deviations from a one-dimensional
   approximation make a negligibly small contribution to the
   total luminosity; this allows disregarding them in the first
   approximation.

In the quasistationary regime, the temperature of a
   neutron star increases monotonically from the external
   layers of the atmosphere to the interior of the envelope until
   it reaches equilibrium (usually in the outer crust). However,
   magnetars must have sources of heating in the envelopes
   capable of maintaining their high luminosity; for this reason,
   temperatures profiles in magnetar envelopes are 
   nonmonotonic
\cite{Kaminker_ea09}.

\subsection{Cooling curves}

The nonstationary problem is described by the same thermal
   balance equations \cite{PCY07}, but the difference
$
   Q=Q_\nu-Q_{h}$
is
   supplemented by the term
$C\mathrm{e}^{-\Phi}\,\partial T
   /
   \partial t,
$
where $C$ is the
   heat capacity per unit volume \cite{GYP}. 
Strictly speaking, $Q$ should be additionally supplemented by yet
another term describing the release of
   latent melting heat during the movement of the Coulomb
   fluid--crystal interface with a change in temperature. But this
   term is always neglected in the available programs for
   calculating the thermal evolution of neutron stars. Following
   the classical work  \cite{GPE}, the nonstationary problem is solved in
   the interior of a neutron star where density surpasses a certain
   threshold value $\rhob$,
while for external envelopes at $\rho<\rhob$, whose relaxation time is
short relative to characteristic times of
   thermal evolution, a stationary system of equations is solved.
Traditionally following
\cite{GPE}, one chooses
$\rhob=10^{10}$ \gcc, but sometimes other $\rhob$
values prove
   more suitable, depending on the concrete problem of interest
\cite{refroid,transie,PCY07,Kaminker_ea09}.
The relation between the heat flow 
   across the boundary $\rhob$  and the temperature $\Tb$ at this boundary
   obtained by solving the stationary problem for the envelopes
   serves as a boundary condition for the nonstationary problem
   in the internal region. It primarily depends on heat 
   conductivity in the \emph{sensitivity strip}
on the $\rho-T$ plane near the ``\emph{turning point},'' where
$\kappa_{e}\sim\kappa_{r}$ \cite{GPE}.
Analytic estimates for
   the position of this point were obtained in \cite{VP01}.

\begin{figure}[t]
\begin{center}
\includegraphics[width=\columnwidth]{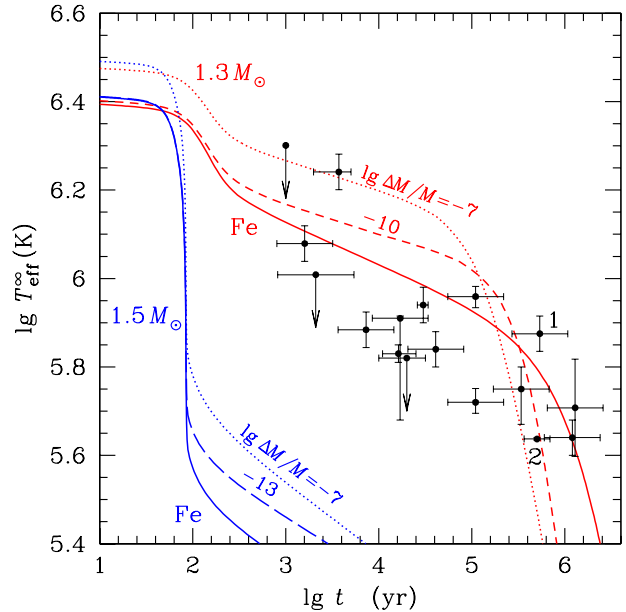}
\caption{Cooling curves of neutron stars compared with some 
   observation-based estimates of their temperatures and ages \cite{Yak08}. 
The cooling
   curves for different models of chemical composition of the heat-insulating
   envelope (in accordance with \cite{refroid}) correspond to different accumulated
   masses
$\Delta M$: of light elements: solid and dotted curves correspond to an iron
   envelope and an envelope of lighter nuclear composition, dashed curves
   correspond to a partly substituted envelope. The upper three curves
   correspond to a star with the mass $M=1\dec3\,M_\odot$, 
undergoing standard
   cooling by Murca processes and the lower three, to a star with the mass $M=1\dec5\,M_\odot$, 
undergoing enhanced cooling by direct Urca processes. Dots
   with error bars correspond to the published estimates of ages $t$ and
   effective surface temperatures
$\Teff^\infty$ of neutron stars; arrows pointing
   down indicate the upper bounds on $\Teff^\infty$.
\label{fig:accr}}
\end{center}
\end{figure}
\begin{figure}[t]
\begin{center}
\includegraphics[width=\columnwidth]{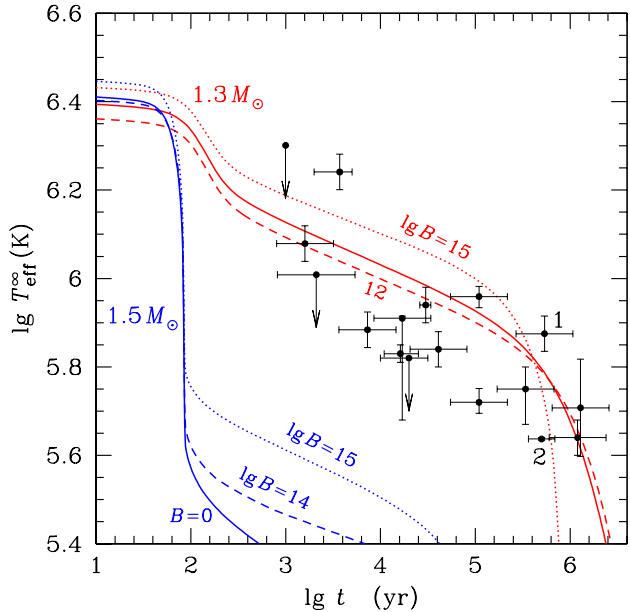}
\caption{The same as in Fig.~\ref{fig:accr}, but for
the iron envelope and different magnetic fields: 
solid and dotted curves correspond to
$B=0$ and $10^{15}$ G respectively, while the dashed curves
correspond to an intermediate case.
\label{fig:magn}}
\end{center}
\end{figure}

The quantities  $\Ts$, $\Teff$, and $L_\gamma$ are defined in the local
reference frame at the neutron star surface. The ``apparent''
quantities in the frame of a distant observer
should be corrected for the redshift
(see Sect.~\ref{sect:GR}).
The solution of the cooling
   problem is described at greater depth in \cite{GYP}
(see also reviews \cite{YakovlevPethick,Yak08} and references therein).

The envelopes of a neutron star at birth consist of iron-group elements, which explains why calculations of cooling
   were for a long time made for iron-rich shells alone. However,
   the envelopes of a star that has passed through an accretion
   stage may consist of lighter elements. Accreted envelopes
   have a higher electron conductivity than iron-rich ones
   because more weakly charged ions less effectively scatter
   electrons. In other words, accretion
   makes the envelopes more ``transparent'' to the passing heat
\cite{PCY97}. The core temperature at the neutrino cooling stage is
   regulated by neutrino emission and is practically independent
   of the properties of the envelopes; therefore, their 
   transparency makes the star brighter due to an enhanced
$\Teff$. At the
   later photon stage, the transparent envelopes more readily
   transmit heat and the star fades away faster. These effects are
   especially demonstrative when comparing the solid, dashed,
   and dotted curves\footnote{The cooling curves presented in
   Figs.~\ref{fig:accr} and \ref{fig:magn}
were calculated by
   D G Yakovlev for Ref.~\cite{CSP06} using a relatively soft
   equation of state, for which the direct Urca processes
   open at $M>1\dec462\,M_\odot$
(see \cite{YakovlevPethick}).}  in Fig.~\ref{fig:accr}, the difference between which is
   due to the different mass of accreted matter $\Delta M$. 

Similarly, cooling depends on a superstrong magnetic
   field. In a strong magnetic field where the electron cyclotron
   frequency w~ exceeds the typical frequency of their collisions
   with plasma ions, heat transfer across magnetic field lines is
   hampered; therefore, those regions in which the lines are
   directed toward the surface become cooler. Oscillations of the
   heat conductivity coefficients (see Fig.~\ref{fig:elounda}), caused by Landau
   quantization facilitate heat transfer along magnetic force
   lines on an average, making the regions near the magnetic
   poles hotter. Taken together, enhanced luminosity near the
   poles and its reduction at the equator make integral
   luminosity of a star in a moderate magnetic field
$B\lesssim10^{13}$~G
   virtually the same as in the absence of a magnetic field.
   Although the temperature distribution over the surface
   depends on the field strength and configuration, the integral
   luminosity is virtually unrelated to $B$ for a moderate dipole field
   \cite{refroid} as well as for a moderate small-scale
   magnetic field \cite{PUC}. However,
   in the superstrong field of  magnetars, $B\gtrsim10^{14}$~G,
   the enhancement of the transparency near the poles
   is more significant, which causes the effect
   of the integral transparency enhancement
   of the envelopes,
   analogous to the effect of accretion, as shown in Fig.~\ref{fig:magn}.

\subsection{Effective temperatures}

The upper and lower groups of three curves in each of Figs.~\ref{fig:accr}
and \ref{fig:magn} correspond to standard and enhanced cooling. The latter
   occurs when direct Urca processes operate at the neutrino
   cooling stage in a star of a sufficiently large mass. The dots
   with error bars indicate estimated ages $t$ and effective surface
   temperatures $\Teff^\infty$, obtained from observational data
   summarized in Ref.~\cite{Yak08}. We see that under
   favorable conditions, cooling curves give an idea of the stellar
   mass and properties of envelopes: the coldest stars of a given
   age appear to undergo enhanced cooling and are therefore
   massive, whereas the hottest ones have accreted envelopes. 

Enhanced cooling may be a consequence not only of direct
   Urca processes in $npe\mu$-matter but also of analogous hyperon
   and quark Urca processes in exotic models of the inner core of
   a neutron star \cite{Haensel95}. The rate of direct Urca processes is
   limited by a gap in the energy spectrum of superfluid nucleons
    \cite{Yak-super}; therefore, nucleon superfluidity smooths the dependence
   of cooling curves on stellar mass, making the ``weighing'' of the
   star by measuring its effective temperature more feasible \cite{Yak08}.
Moreover, nucleon superfluidity, as well as hyperon
   and quark (color) superfluidity in exotic models, decreases the
   heat capacity of the stellar core \cite{Yak-super}, which also affects
   cooling
\cite{Yak08}.

Thus, comparison of measured ages and
   temperatures with cooling curves allows one, in principle, to
   determine stellar
   mass, or to conjecture the composition of the core and 
   heat-insulating envelopes when the mass is known from 
   independent estimates. But the results thus obtained should be
   interpreted with caution for the following reasons. Effective
   temperatures $\Teff^\infty$ are typically measured by varying the
   parameters of a theoretical model used to calculate the
   emission spectrum. The parameters are chosen so as to most
   accurately reproduce the observed spectrum, but the final
   result strongly depends on the choice of the model. Certain
   estimates $\Teff^\infty$,
presented in Figs.~\ref{fig:accr} and \ref{fig:magn},
were obtained using
   models of nonmagnetic and magnetic hydrogen atmospheres,
   while others simply assume that radiation is described by the
   Planck spectrum. For example, the fit to the three-component
   model spectrum was used for the thermal component of the PSR B1055--52 spectrum: the power-law component was
   added to two black-body components, whose cooling is
   believed to be responsible for heat radiation from the surface \cite{PZ03}.
The result of this fitting is marked in the figures by the
   numeral 1, while the one marked by 2 was obtained for the
   radio-quiet neutron star RX J1856.4--3754 based on a physical model
   of the magnetic atmosphere \cite{Ho_ea07} 
 (we will discuss it in Sect.~\ref{sect:RM} in more detail). 
 For comparison, the authors of
\cite{Ho_ea07} also
fitted the
   observed X-ray spectrum of RX J1856.4--3754 
with the Planck spectrum. When plotted in Figs.~\ref{fig:accr} and  \ref{fig:magn}, 
the result of such a simplistic fitting would almost coincide with 
the point 1. This means that the
   systematic error in the position of point 1 resulting from the
   absence of a model for the formation of the PSR B1055--52 spectrum may be of the same order as the distance between
   points 1 and 2, that is, significantly greater than the statistical
   fitting error.

\subsection{Masses and radii\label{sect:RM}}

Analysis of the thermal radiation spectrum of a neutron star
   provides information not only on its effective temperature but
   also on the radius and mass. Let us first consider the Planck
   spectrum disregarding interstellar absorption and possible
   non uniformity of the temperature distribution over the
   surface. The position of the spectral maximum gives
$\Teff^\infty$, and the total (bolometric) flux of incoming
   radiation
$F_\mathrm{bol}^\infty$ is found from its measured intensity. If a star is at a
   distance $D$, its apparent photon luminosity is obtained as
$L_\gamma^\infty=4\pi D^2 F_\mathrm{bol}^\infty$. 
On the other hand, according tot he Stefan--Boltzmann law
$L_\gamma^\infty=4\pi\sigma_\mathrm{SB}\,R_\infty^2
\,(\Teff^\infty)^4$, which allows one to estimate $R_\infty$.

In reality, the comparison of theoretical and measured spectra
   actually involves more unknown variables. The spectrum is
   distorted by absorption in the interstellar gas; therefore,
   spectral analysis can be used to determine average gas
   concentration over the line of sight. If the distance $D$ is
   unknown, it can be estimated under the assumption of the
   typical concentration of the interstellar gas in a given galactic
   region, using $D$ as a fitting parameter. The temperature
   distribution over the stellar surface may be nonuniform. For
   example, the thermal spectrum in the presence of hot polar
   caps consists of two blackbody components,
   each
   having its own values of $\Teff^\infty$ and $R_\infty$.
    Finally, because the star is not a perfect
   black body, the real radiation spectrum differs from the
   Planck spectrum. Spectrum simulation is a difficult task,
   involving solution of the equations of hydrostatic 
   equilibrium, energy balance, and radiation transmission
   \cite{Mihalas}. Coefficients of these equations depend on the chemical
   composition of the atmosphere, effective temperature, 
   acceleration of gravity, and magnetic field. Different assumptions
   of the chemical composition, 
$\Teff$, $z_g$, and $B$, values lead to
   different model spectra; their comparison with the observed
   spectrum permits obtaining acceptable parameter values.
   Knowing the shape of the spectrum, allows calculating the
   fitting coefficient by the Stefan--Boltzmann formula and
   finding $R_\infty$ from $F_\mathrm{bol}^\infty$.
The simultaneous finding of $z_g$ and $R_\infty=R(1+z_g)$ allows one to calculate the mass $M$
on the base of Eqs.~(\ref{r_g}) and (\ref{z_g}).

Let us consider some of the problems arising from the estimation
   of parameters of neutron stars from their observed thermal
   spectra using the ``Walter star''  RX J1856.4--3754 as an example.
   It is a nearby radio-quiet neutron star, discovered in 1996 
as a soft X-ray source \cite{Walter96} and identified a year later in
the optical range \cite{Walter97}. The
   first parallax measurement by the ``planetary camera'' (PC) 
   on board the Hubble space observatory gave the value $D\approx60$ pc
\cite{Walter01}, which corresponded to a very small radius $R$.
A later improvement of the distance together with a preliminary spectral
analysis gave a somewhat larger $R$ which might correspond to a quark
star
\cite{Drake}. A subsequent reanalysis of the data gave $D\approx120$ pc
\cite{WL02}, and independent 
   treatment of the same data by different authors has led to
   $D\approx140$ pc \cite{Kaplan02}. Finally, the measurement of parallax by the
   high resolution camera (HRC) of the same observatory
   yielded $D\approx160$ pc.  At the same time, it turned out that
the spectrum of the Walter star is not described by the 
   blackbody model: the fit of its X-ray region to the Planck spectrum
   predicts a much weaker luminosity in the optical range than
   the observed one. Attempts to describe the observed spectrum
   in terms of the models of atmospheres of different chemical
   compositions without a magnetic field and of the two-component
    Planck model are reported in \cite{Pons} and \cite{Burwitz},
    respectively.
It
   turned out that the hydrogen atmosphere model reproducing
   the X-ray spectral region predicts too high a luminosity in the
   optical range and models of atmospheres of a different
   chemical composition predict absorption lines unseen in
   observational data. Fitting to the two-component model
   for the soft component of the composite spectrum leads to a
   bound on the radius
$R_\infty>17$ km $\times(D/120$ pc),  
which is
   difficult to relate to theoretical calculations of neutron star
   radii.

A simulation of the neutron star spectrum based on the
   solution of a system of equations for radiation transfer in a
   partly ionized hydrogen atmosphere of finite thickness above
   the condensed surface in a strong magnetic field was proposed
   in \cite{Ho_ea07}. The authors used the atmosphere model from
   \cite{KK}, based on the equation of state of the hydrogen plasma in a
   strong magnetic field and absorption/scattering coefficients
   in such a plasma presented in \cite{PC03}. At
$B\sim(3$\,--\,$4)\times10^{12}$~G,
$\Teff^\infty=(4\dec34\pm0\dec03)\times10^5$~K,
$z_g=0\dec25\pm0\dec05$, and
$R_\infty=17\dec2^{+0\dec5}_{-0\dec1}\,d_{140}$ km, they managed for the first time to
   reproduce the measured spectrum of RX J1856.4--3754 in
   the X-ray to optical frequency range within the measurement
   errors of the best space and terrestrial observatories. Here, the
   errors are given at the significance levels $1\sigma$, and $d_{140}\equiv
D/(140$ pc). Taking relations
(\ref{r_g})\,--\,(\ref{z_g}) into account,
one founds from these estimates, that for this neutron star
$R=13\dec8^{+0\dec9}_{-0\dec6}\,d_{140}$ km and
$M=1\dec68^{+0\dec22}_{-0\dec15}\,d_{140}\,M_\odot$. Forgetting for a
moment the multiplier $d_{140}$, one might conclude that the 68\% 
confidence area lies above all theoretical dependences $R(M)$, 
shown in Fig.~\ref{fig:RMpsr}. It could mean that
either the measurements
or the theoretical model are not accurate (if not to consider
a possibility of a superstiff equation of state,
not shown in Fig.~\ref{fig:RMpsr}). 
The estimate $D\approx 160$ pc \cite{vKK07}
shifts the values of $R$ and $M$ still farther from the theoretical 
dependences $R(M)$. Moreover, such a massive star should have
   undergone enhanced cooling, which is not observed in Figs.~\ref{fig:accr}
   and \ref{fig:magn}.
However, confirmation of the estimate $D\approx120$ pc
\cite{WL02} in a recent paper
\cite{Walter10} eliminates these
   contradictions. 

Thus, the uncertainty of distance measurements
   proves more important than the inaccuracy of spectrum
   fitting. An even greater uncertainty is associated with the
   choice of a theoretical model. For example, fitting the X-ray
   region of the RX
J1856.4--3754 spectrum with the black-body
   spectrum, presented in \cite{Ho_ea07}
for means of comparison, gives
$R_\infty\approx5\,d_{140}$ km. 

Similar problems are encountered in the analysis of all
   known thermal spectra of isolated neutron stars. They are not
   infrequently supplemented by uncertainties of spectrum
   division into thermal and nonthermal components (see \cite{Zavlin09} 
   and the references therein).

\section{Conclusions\label{sect:concl}}

Neutron stars are miraculous objects in which Nature
   assembled its puzzles, whose solution is sought by seemingly
   unrelated branches of science, such as the physics of outer
   space and the micro-world, giant gravitating masses, and
   particle particles. This makes neutron stars unique cosmic
   laboratories for the verification of basic physical concepts. In
   the last 50 years, both theoretical and observational studies of
   neutron stars have been developing at a progressively faster
   pace following advances in nuclear and elementary physics on
   the one hand, and astronomy and experimental physics on the
   other hand.

The present review outlines some aspects of neutron star
   physics, describes methods for measuring their temperature,
   masses, and radii, and illustrates the relation between the
   theoretical interpretation of these data and the solution of
   fundamental physical problems.

The work was supported by the Russian Foundation for
   Basic Research (grant 08-02-00837) and the State
   Program for Support of Leading Scientific Schools of the
   Russian Federation (grant NSh-3769.2010.2).\\\\

\noindent{\bf\textit{Note added in proof}}\\
When this paper was being prepared for publication,
   P Demorest et al. reported the record-breaking mass
   $M=1.97\pm0.04,M_\odot$ of the neutron star in the PSR J1614--2230 
   binary system \cite{Demorest}. This estimate
   was obtained by measuring the Shapiro delay parameters (see
   Section 2.1 of the present review). Plotting it in our Fig.~4
   results in appearance of 
   the corresponding vertical strip slightly to
   the left of point \textit{2}. In conformity with the discussion in
   Section 4.4, it prompts that superdense
   matter cannot be characterized by equations of state softer than Sly.

\newcommand{\artref}[2]{\textit{#1} \textbf{#2}}
\newcommand{\AandA}[1]{\artref{Astron.\ Astrophys.}{#1}}
\newcommand{\ApJ}[1]{\artref{Astrophys.\ J.}{#1}}
\newcommand{\ApSS}[1]{\artref{Astrophys.\ Space Sci.}{#1}}
\newcommand{\MNRAS}[1]{\artref{Mon.\ Not.\ R.\ astron.\ Soc.}{#1}}
\newcommand{\PRL}[1]{\artref{Phys.\ Rev.\ Lett.}{#1}}
\newcommand{\ARAA}[1]{\artref{Annu.\ Rev.\ Astron.\ Astrophys.}{#1}}

\pagebreak

\renewcommand{\refname}{References}
\makeatletter
\renewcommand\@biblabel[1]{#1.}
\makeatother

\label{sect:bib}


\begin{thebibliography}{99}
\itemsep=4pt

\bibitem{Fortov}
Fortov V E
\artref{Phys. Usp.}{52}, 615 (2009)

\bibitem{Glendenning}
Glendenning N K
\textit{Compact Stars: Nuclear Physics, Particle
Physics, and General Relativity}
(2nd ed.) (New York: Springer, 2000)

\bibitem{Shklovsky-book}
Shklovskii I S
\textit{Stars: Their Birth, Life, and Death} (San Francisco:
Freeman, 1978)

\bibitem{Ginzburg64}
Ginzburg V L
\artref{Sov.\ Phys.\ Doklady}{9}, 329 (1964)

\bibitem{GinzburgOzernoy}
Ginzburg V L, Ozernoi L M 
      \artref{Sov. Phys. JETP}{20}, 689 (1965)

\bibitem{GinzburgKirzhnits}
Ginzburg V L, Kirzhnits D A
\artref{Sov. Phys. JETP}{20}, 1346 (1965)

\bibitem{Migdal59}
Migdal A B
\artref{Sov. Phys. JETP}{10}, 176 (1960)

\bibitem{Ginzburg69sup}
Ginzburg V L
\artref{Sov. Phys. Usp.}{12}, 241 (1969)

\bibitem{GinzburgSyr}
Ginzburg V L, Syrovatskii S I
\artref{Sov.\ Phys.\ Doklady}{9}, 831 (1965)

\bibitem{GinzburgKirzhniz}
Ginzburg V L, Kirzhnits D A
\artref{Nature}{220}, 148 (1968)

\bibitem{GinzburgZZ68}
Ginzburg V L, Zheleznyakov V V, Zaitsev V V
\artref{Nature}{220}, 355 (1968)

\bibitem{GinzburgZ69}
Ginzburg V L, Zaitsev V V
\artref{Nature}{222}, 230 (1969)

\bibitem{GinzburgZZ69}
Ginzburg V L, Zheleznyakov V V, Zaitsev V V
\ApSS{4}, 464 (1969)

\bibitem{GinzburgZ75}
Ginzburg V L, Zheleznyakov V V
\ARAA{13}, 511 (1975)

\bibitem{GinzburgZ70a}
Ginzburg V L, Zheleznyakov V V
\artref{Comments on Astrophysics and Space Physics}{2}, 167 (1970)

\bibitem{GinzburgZ70b}
Ginzburg V L, Zheleznyakov V V
\artref{Comments on Astrophysics and Space Physics}{2}, 197 (1970)

\bibitem{Ginzburg71hi}
Ginzburg V L
\artref{Highlights of Astronomy}{2}, 737 (1971)

\bibitem{GinzburgUsov}
Ginzburg V L, Usov V V
\artref{JETP Lett.}{15}, 196 (1972)

\bibitem{Ginzburg71}
Ginzburg V L
\artref{Sov.\ Phys.\ Usp.}{14}, 83 (1971);
\artref{Highlights of Astronomy}{2}, 20 (1971)

\bibitem{ATNF}
\textit{ATNF Pulsar Catalogue}
(http://www.atnf.csiro.au/research/pulsar/psrcat/);
Manchester R N et al.
\artref{Astron.\ J.}{129}, 1993 (2005)

\bibitem{Bisno06}
Bisnovatyi-Kogan G S
\artref{Phys.\ Usp.}{49}, 53 (2006)

\bibitem{Mereghetti08}
Mereghetti S
\artref{Astron.\ Astrophys.\ Rev.}{15}, 225 (2008)

\bibitem{Ginzburg}
Ginzburg V L
\textit{The Propagation of Electromagnetic Waves in Plasmas}
(2nd ed.) (London: Pergamon, 1970)

\bibitem{ShapiroTeukolski}
Shapiro S L, Teukolsky S A
\textit{Black Holes, White Dwarfs, and Neutron Stars:
The Physics of Compact Objects} 
(New York: Wiley, 1983)

\bibitem{NSB1}
Haensel P, Potekhin A~Y, Yakovlev D~G
\textit{Neutron Stars 1: Equation of State and Structure}
(New York: Springer, 2007)

\bibitem{Yak-super}
Yakovlev D G, Levenfish K P, Shibanov Yu A
\artref{Phys.\ Usp.}{42}, 737 (1999)

\bibitem{Beskin99}
Beskin V S
\artref{Phys.\ Usp.}{42}, 1071 (1999)

\bibitem{Malov04}
Malov I F  (Moscow: Nauka, 2004)
\textit{Radiopul'sary (Radiopulsars)}
(Moscow: Nauka, 2004) [in Russian]

\bibitem{Michel}
Michel F C
\artref{Adv.\ Space Res.}{33}, 542 (2004)

\bibitem{Yak-neu}
Yakovlev D G et al.
 \artref{Phys.\ Rep.}{354}, 1 (2001)

\bibitem{MisnerTW}
Misner C~W, Thorne K~S, Wheeler J~A
\textit{Gravitation}
(San Francisco: Freeman \& Co., 1973)

\bibitem{Tolman39}
Tolman R C
\artref{Phys.\ Rev.}{55}, 364 (1939)

\bibitem{Oppenheimer39}
Oppenheimer J R, Volkoff G M,
\artref{Phys.\ Rev.}{55}, 374 (1939)

\bibitem{Hessels_ea06}
Hessels J W T et al.
\artref{Science}{311}, 1901 (2006)

\bibitem{BonazzolaG}
Bonazzola S, Gourgoulhon E,
\AandA{312}, 675 (1996)

\bibitem{Brag00}
Braginskii V B
\artref{Phys.\ Usp.}{43}, 691 (2000)

\bibitem{PoPro07}
Popov S B, Prokhorov M E
\artref{Phys.\ Usp.}{50}, 1123 (2007)

\bibitem{KramerStairs}
Kramer M, Stairs I H,
\ARAA{46}, 541 (2008)

\bibitem{Istomin91}
Istomin Ya B
\artref{Sov.\ Astron.\ Lett.}{17}, 301 (1991)

\bibitem{Kramer98}
Kramer M
\ApJ{509}, 856 (1998)

\bibitem{Champion}
Champion D J et al.
\artref{Science}{320}, 1309

\bibitem{Freire}
Freire P,
arXiv:0907.3219

\bibitem{Shklovsky-SNe}
Shklovskii I S
\textit{Supernovae} (New York:
Wiley, 1968)

\bibitem{ImshennikNadyozhin}
Imshennik V S, Nadyozhin D K, 
\artref{Usp. Fiz. Nauk}{156}, 561 (1988)
[English transl.: 
in \textit{Soviet 
Scientific Reviews, 
Ser. E: Astrophys. and Space Phys.}, Vol.\,\textbf{7}
 (Amsterdam: Harwood), 75 (1989)]

\bibitem{Imshennik95}
Imshennik V S
\artref{Space Sci.\ Rev.}{74}, 325 (1995)

\bibitem{Arnett}
Arnett D
\textit{Supernovae and Nucleosynthesis}
 (Princeton: Princeton University Press, 1996)

\bibitem{WoosleyJanka}
Woosley S, Janka H-T
\artref{Nature Physics}{1}, 147 (2005)

\bibitem{Paczynski98}
Paczy\'nski B
\ApJ{42}, 145 (1998)

\bibitem{Postnov99}
Postnov K A
\artref{Phys.\ Usp.}{42}, 469 (1999)

\bibitem{StrohmayerBildsten}
Strohmayer T E, Bildsten L
in \textit{Compact Stellar X-Ray Sources}
(Eds.\ W H G Lewin, M van der Klis)
(Cambridge: Cambridge University Press, 2006), p.~113

\bibitem{Ruderman69}
Ruderman M
\artref{Nature}{223} 597 (1969)

\bibitem{BP71}
Baym G, Pines D
\artref{Ann.\ Phys.\ (N.Y.)}{66}, 816 (1971)

\bibitem{Blaes-ea89}
Blaes O et al.
\ApJ{343}, 839--848 (1989).

\bibitem{HDP90}
Haensel P, Denissov A, Popov S
\AandA{240}, 78 (1990)

\bibitem{Alpar98}
Alpar M A
\artref{Adv.\ Space Res.}{21}, 159 (1998)

\bibitem{Franco-ea}
Franco L M, Bennett L, Epstein R I
\ApJ{543}, 987 (2004)

\bibitem{Reisen-roto}
Reisenegger A et al.
\ApJ{653}, 568 (2006)

\bibitem{BisnoChech}
Bisnovatyi-Kogan G S, Chechetkin V M
\artref{Sov.\ Phys.\ Usp.}{22}, 89 (1979)

\bibitem{HZ90}
Haensel P, Zdunik J L
\AandA{227}, 431 (1990)

\bibitem{HZ03}
Haensel P, Zdunik J L
\AandA{404}, L33 (2003)

\bibitem{Lipunov}
Lipunov V~M
\textit{Astrophysics of Neutron Stars}
(Berlin: Springer, 1992)

\bibitem{BaadeZwicky}
Baade W, Zwicky F
\artref{Phys.\ Rev.}{45}, 138 (1934)

\bibitem{Chadwick32}
Chadwick J
\textit{Nature} \textbf{129}, 312 (1932)

\bibitem{Rosenfeld74}
Rosenfeld L, 
in \textit{Astrophysics \& Gravitation},
Proc.\ 16th Solvay Conf.\ on Physics 
(Brussels: Universit\'e de Bruxelles, 1974), p.~174

\bibitem{Landau32}
Landau L D
\textit{Phys.\ Z.\ Sowjetunion}
\textbf{1}, 285 (1932)

\bibitem{Zwicky38}
Zwicky F
\ApJ{88}, 522 (1938)

\bibitem{Cameron59}
Cameron A G W
\ApJ{130}, 916 (1959)

\bibitem{BMP58}
Bohr A, Mottelson B R, Pines D
\artref{Phys.\ Rev.}{110}, 936 (1958)

\bibitem{AS60}
Ambartsumyan V A, Saakyan G S
\artref{Sov.\ Astron.}{4}, 187 (1960)

\bibitem{Zeld61}
Zeldovich Ya B
\artref{Sov.\ Phys.--JETP}{14}, 1143 (1962)

\bibitem[Bahcall \& Wolf(1965a)]{BahcallWolf}
Bahcall J N, Wolf R A
\artref{Phys.\ Rev.}{140}, B1452 (1965)

\bibitem[Chiu \& Salpeter(1964)]{ChiuSalpeter64}
Chiu H-Y, Salpeter E E
\PRL{12}, 413 (1964)

\bibitem[Stabler(1960)]{Stabler60}
Stabler R
\textit{Ph.D.\ Thesis}  (Ithaca, NY: Cornell University, 1960)

\bibitem[Morton(1964)]{Morton64}
Morton D C
\artref{Nature}{201}, 1308 (1964)

\bibitem[Bahcall \& Wolf(1965b)]{BahcallWolfb}
Bahcall J N, Wolf R A
\ApJ{142}, 1254 (1965)

\bibitem[Tsuruta \& Cameron(1966a)]{tc66}
Tsuruta S, Cameron A G W
\artref{Canadian J.\ Phys.}{44}, 1863 (1966)

\bibitem[Pacini(1967)]{Pacini67}
Pacini F
\artref{Nature}{216}, 567 (1967)

\bibitem[Giacconi \etal(1962)]{Giacconi62}
Giacconi R et al.
\PRL{9}, 439 (1962)

\bibitem{BeckerTrumper}
Becker W, Tr\"umper J
\AandA{326}, 682 (1997) 

\bibitem[Zeldovich \& Guseynov(1966)]{zg66}
Zeldovich Ya B, Guseynov O H
\ApJ{144}, 840 (1966)

\bibitem{Kardashev}
Kardashev N S
\textit{Sov.\ Astron.} \textbf{8}, 643 (1965)

\bibitem{Hewish68}
Hewish A et al.
\artref{Nature}{217}, 709 (1968)

\bibitem{Hewish75}
Hewish A
\artref{Rev.\ Mod.\ Phys.}{47}, 567 (1975)

\bibitem{Gold68}
Gold T
\artref{Nature}{218}, 731 (1968)

\bibitem{Shklovsky67}
Shklovsky I S
\ApJ{148}, L1 (1967)

\bibitem{deFreitas77}
de Freitas Pacheco J A, Steiner J E, Neto A D
\AandA{55}, 111 (1977)

\bibitem{CameronScoX1}
Cameron A G W
\artref{Nature}{215}, 464 (1967)

\bibitem[Yakovlev \& Pethick(2003)]{YakovlevPethick}
Yakovlev D G, Pethick C J
\artref{Annu.\ Rev.\ Astron.\ Astrophys.}{42}, 169 (2004)

\bibitem[Zavlin(2009)]{Zavlin09}
Zavlin V E
in 
\textit{Neutron Stars and Pulsars}
(Ed W~Becker) (New York: Springer, 2009), p.~181
  
\bibitem[Walter \etal(1996)]{Walter96}
Walter F M, Wolk S J, Neuh\"auser R
\artref{Nature}{379}, 233 (1996)

\bibitem[de Luca(2008)]{DeLuca08}
 de Luca A
\artref{AIP Conf.\ Proc.}{983}, 311 (2008)

\bibitem[Haberl(2007)]{Haberl07}
Haberl F
\ApSS{308}, 181 (2007)

\bibitem{HalpernGotthelf}
Halpern J P, Gotthelf E V
\ApJ{709}, 436 (2010)

\bibitem{PopovProkhorov}
Popov S B, Prokhorov M E
\textit{Astrofizika Odinochnykh Neitronnykh
      Zvezd: Radiotikhie Neitronnye Zvezdy i Magnitary (Astrophysics of
      Isolated Neutron Stars: Radio-Quiet Neutron Stars and Magnetars)}
      (Moscow: GAISh MGU, 2002)

\bibitem{PC03}
Potekhin A Y, Chabrier G
\ApJ{585}, 955 (2003)

\bibitem{P10}
Potekhin A Y
\AandA{518}, A24 (2010)

\bibitem[Thompson(2001)]{Thompson01}
Thompson C,
 in \textit{The Neutron Star -- Black Hole Connection}
(Eds C~Kouveliotou, J~Ventura, E~Van den Heuvel)
 (Dordrecht: Kluwer, 2001), p.~369

\bibitem[Urpin \& Konenkov(2008)]{UrpinKonenkov08}
Urpin V, Konenkov D
\AandA{483}, 223 (2008)

\bibitem[Pons \etal(2009)Pons, Miralles, \& Geppert]{Pons_MG}
Pons J A, Miralles J A, Geppert U
\AandA{496}, 207 (2009)\

\bibitem[Kaminker \etal(2009)]{Kaminker_ea09}
Kaminker A D et al.
\MNRAS{395}, 2257 (2009)

\bibitem{ManchesterTaylor}
Manchester R, Taylor J
\textit{Pulsars}
(San Francisco: Freeman, 1977)

\bibitem{GreenStephenson}
Green D A, Stephenson F R
\textit{Historical Supernovae}
(Oxford: Clarendon Press, 2002)

\bibitem[Zavlin(2007)]{Zavlin07}
 Zavlin V E
\ApSS{308}, 297 (2007)

\bibitem{MalovMachabeli09}
Malov I F, Machabeli G Z  
\textit{Anomal'nye Pul'sary (Anomalous
      Pulsars)}
(Moscow: Nauka, 2009)

\bibitem[Ertan \etal(2007)]{ErtanAXP}
Ertan \"U et al.
\ApSS{308}, 73 (2007)

\bibitem[van der Klis \etal(1985)]{vanderklisetal85}
van der Klis M et al.
\artref{Nature}{316}, 225 (1985)

\bibitem[van der Klis \etal(2000)]{vanderklis00}
van der Klis M
\artref{Annu.\ Rev.\ Astron.\ Astrophys.}{38}, 717 (2000)

\bibitem{Kluz07}
Klu\'zniak M et al.
\artref{Rev.\ Mex.\ Astron.\ Astrophys.}{27}, 18 (2007)

\bibitem{Tagger07}
Tagger M
\artref{Rev.\ Mex.\ Astron.\ Astrophys.}{27}, 26 (2007)

\bibitem[Shaposhnikov \& Titarchuk(2004)]{ShaposhnikovTitarchuk}
Shaposhnikov N, Titarchuk L
\ApJ{606}, L57 (2004)

\bibitem{Steiner}
Steiner A W, Lattimer J M, Brown E F,
\ApJ{722}, 33 (2010)

\bibitem[Brown \etal(1998)Brown, Bildsten, \& Rutledge]{Brown_BR98}
Brown E F, Bildsten L, Rutledge R E
\ApJ{504}, L95 (1998)

\bibitem[Brown \& Cumming(2009)]{BrownCumming09}
Brown E F, Cumming A
\ApJ{698}, 1020 (2009)

\bibitem[Shternin \etal(2007)]{Shternin07}
Shternin P S et al.
\MNRAS{382}, L43 (2007)

\bibitem[Akmal \etal(1998)]{APR}
Akmal A, Pandharipande V R, Ravenhall D G
\artref{Phys.\ Rev. C}{58}, 1804 (1998)

\bibitem[Douchin \& Haensel(2001)]{SLy}
Douchin F, Haensel P
\AandA{380}, 151 (2001)

\bibitem[Pandharipande \& Ravenhall(1989)]{FPS}
Pandharipande V R, Ravenhall D G,
in \textit{Nuclear Matter and Heavy Ion
Collisions}
(Eds. M Soyeur et al.) (Dordrecht: Reidel, 1989), 
p.~103

\bibitem[Douchin \& Haensel(2000)]{SLy-edge}
Douchin F, Haensel P
\artref{Phys.\ Lett. B}{485}, 107 (2000)

\bibitem[Heiselberg \& Hjorth-Jensen(2000)]{HeiselHj00}
Heiselberg H, Hjorth-Jensen M
\textit{Phys.\ Rep.} \textbf{328}, 237 (2000)

\bibitem{HP04}
Haensel P, Potekhin A Y
\AandA{428}, 191 (2004)

\bibitem[Bocquet \etal(1995)]{Bocquet}
Bocquet M et al. 
 \AandA{301}, 757 (1995)

\bibitem[Walecka(1974)]{Walecka}
Walecka J D
\artref{Ann.\ Phys.\ (N.Y.)}{83}, 491 (1974)

\bibitem{GamowS}
Gamow G, Schoenberg M
\artref{Phys.\ Rev.}{59}, 539 (1941)

\bibitem{Gamow}
Gamow G
\textit{My World Line: An Informal Autobiography}
(New York: Viking Press, 1970)

\bibitem{Haensel95}
Haensel P
\artref{Space Sci.\ Rev.}{74}, 427 (1995)

\bibitem[Salpeter(1960)]{Salpeter60}
Salpeter E E
\artref{Ann.\ Phys.\ (N.Y.)}{11}, 393 (1960)

\bibitem[Balberg \& Gal(1997)]{BGN}
Balberg S, Gal A
\artref{Nucl.\ Phys. A}{625}, 435 (1997)

\bibitem{Migdal71}
Migdal A B
\artref{Sov.\ Phys.--JETP}{34}, 1184 (1972)

\bibitem[Sawyer(1972b)]{SawyerPi}
Sawyer R F
\PRL{29}, 382 (1972); erratum: \textit{ibid.}, \textbf{29}, 823
(1972)

\bibitem[Scalapino(1972)]{Scalapino}
Scalapino D J
\PRL{29}, 386 (1972)

\bibitem{Migdal77}
Migdal A B
\artref{Sov.\ Phys. Uspekhi}{20}, 879 (1977)

\bibitem[Kunihiro \etal(1993)]{Kunihiro}
Kunihiro T, Takatsuka T, Tamagaki R
\artref{Prog.\ Theor.\ Phys.\ Suppl.}{112}, 197 (1993)

\bibitem[Kaplan \& Nelson(1986)]{KaplanNelson}
Kaplan D B, Nelson A E
\artref{Phys.\ Lett.\ B}{175}, 57 (1986)

\bibitem[Ramos \etal(2001)]{Ramos}
Ramos A, Schaffner-Bielich J, Wambach J
\artref{Lecture Notes in Phys.}{578}, 175 (2001)

\bibitem[Kolomeitsev \& Voskresensky(2003)]{Kolomeitsev}
Kolomeitsev E E, Voskresensky D N
\artref{Phys.\ Rev. C}{68}, 015803 (2003)

\bibitem[Glendenning \& Schaffner-Bielich(1998)]{GlendSchaff}
Glendenning N K, Schaffner-Bielich J
\PRL{81}, 4564 (1998)

\bibitem{QCD}
Dremin I M, Kaidalov A B
\artref{Phys.\ Usp.}{49}, 263 (2006)

\bibitem[Ivanenko \& Kurdgelaidze(1965)]{Ivanenko65}
Ivanenko D D, Kurdgelaidze D F,
\artref{Astrophysics}{1}, 251 (1965)

\bibitem[Collins \& Perry(1975)]{CollinsPerry}
Collins J C, Perry M J
\PRL{34}, 1353 (1975)

\bibitem{Kurkela}
Kurkela A, Romatschke P, Vuorinen A
\artref{Phys.\ Rev. D}{81}, 105021 (2010)

\bibitem{Blaschke09}
Blaschke D et al.
\artref{Phys.\ Rev. C}{80}, 065807 (2009)

\bibitem{ilios10}
Iosilevskiy I
\artref{Acta Physica Polonica B (Proc.\ Suppl.)}{3}, 589 (2010)

\bibitem[Glendenning(1992)]{Glend92mix}
Glendenning N K
\artref{Phys.\ Rev. D}{46}, 1274 (1992)

\bibitem[Cazzola \etal(1966)]{Cazzola}
Cazzola P, Lucaroni L, Scaringi C
\artref{Nuovo Cimento B}{43}, 250 (1966)

\bibitem[Takemori \& Guyer(1975)]{TakemoriGuyer}
Takemori M T, Guyer R A
\artref{Phys.\ Rev. D}{11}, 2696 (1975)

\bibitem[Pandharipande \& Smith(1975)]{PandhaSmith}
Pandharipande V R, Smith R A
\artref{Phys.\ Lett.\ B}{59}, 15 (1975)

\bibitem[Takatsuka \& Tamagaki(1977)]{TakatsukaTamagaki77}
Takatsuka T, Tamagaki R
\artref{Prog.\ Theor.\ Phys.}{58}, 694 (1977)

\bibitem[Kutschera \& W{\'o}jcik(1989)]{KutscheraWo89}
Kutschera M, W{\'o}jcik W
\artref{Phys.\ Lett. B}{223}, 11 (1989)

\bibitem[Takatsuka \& Tamagaki(1988b)]{TakatsukaTamagaki88b}
Takatsuka T, Tamagaki R
\artref{Prog.\ Theor.\ Phys.}{79}, 274 (1988)

\bibitem[Kristian \etal(1989)]{Kristian89}
Kristian J et al.
\artref{Nature}{338}, 234 (1989)

\bibitem[Kristian(1991)]{Kristian91}
Kristian J
\artref{Nature}{349}, 747 (1991)

\bibitem{PethickRavenhall}
Pethick C J, Ravenhall D G
\artref{Annu.\ Rev.\ Nucl.\ Sci.}{45}, 429 (1995)

\bibitem[Chugunov \& Haensel(2007)]{ChugunovHaensel}
Chugunov A I, Haensel P
\MNRAS{381}, 1143 (2007)

\bibitem[Aguilera \etal(2009)]{Aguilera-neutrons}
Aguilera D N et al.
\PRL{102}, 091101 (2009)

\bibitem[Pethick \& Potekhin(1998)]{PethickPotekhin}
Pethick C J, Potekhin A Y
\artref{Phys.\ Lett. B}{427}, 7 (1998)

\bibitem{RPW83}
Ravenhall D G, Pethick C J, Wilson J R
\PRL{50}, 2066 (1983)

\bibitem{Lorenz93}
Lorenz C P, Ravenhall D G, Pethick C J
\PRL{70}, 379 (1993)

\bibitem[Gusakov \etal(2004)]{Gusakov_ea04}
Gusakov M E et al.
\AandA{421}, 1143 (2004)

\bibitem{PC00}
Potekhin A Y, Chabrier G
\artref{Phys.\ Rev. E}{62}, 8554 (2000)

\bibitem[Lai(2001)]{Lai01}
Lai D
\artref{Rev.\ Mod.\ Phys.}{73}, 629 (2001)

\bibitem{BKPY}
Baiko D A et al.
\PRL{81}, 5556 (1998)

\bibitem[Potekhin \etal(2004)]{KK}
Potekhin A Y et al.
\ApJ{612}, 1034 (2004)

\bibitem[Ho \etal(2008)Ho, Potekhin, \& Chabrier]{HoPC07}
Ho W C G, Potekhin A Y, Chabrier G
\artref{Astrophys,\ J.\ Suppl.\ Ser.}{178}, 102 (2008)

\bibitem[Mori \& Ho(2007)]{MoriHo}
Mori K, Ho W C G
\MNRAS{377}, 905 (2007)

\bibitem[Dall'Osso et al.(2009)Dall'Osso, Shore, \& Stella]{DallOsso_ss}
Dall'Osso S, Shore S N, Stella L
\MNRAS{398}, 1869 (2009)
 
\bibitem[Chandrasekhar \& Fermi(1953)]{ChandraFermi}
Chandrasekhar S, Fermi E
\ApJ{118}, 116 (1953)

\bibitem[Lai \& Shapiro(1991)]{LaiShapiro91}
Lai D, Shapiro E E
\ApJ{383}, 745 (1991)

\bibitem[Reisenegger(2003)]{Reisenegger}
Reisenegger A,
in \textit{Proceedings of the International
Workshop on Strong Magnetic Fields
and Neutron Stars}
(Eds.\ H~J Mosquera Cuesta, H~Per\'ez Rojas, C~A Zen Vasconcellos)
(La Habana, Cuba: ICIMAF, 2003) p.~33

\bibitem[Bisnovatyi-Kogan(1992)]{Bisno92dynamo}
Bisnovatyi-Kogan G S
\ApSS{189}, 147 (1992)

\bibitem[Thompson \& Duncan(1993)]{TD93}
Thompson C, Duncan R C
\ApJ{408}, 194 (1993)

\bibitem[Baym \etal(1969)]{BaymPP}
Baym G, Pethick C, Pines D
\artref{Nature}{224}, 674 (1969)

\bibitem[Cumming \etal(2004)]{Cumming_ea04}
Cumming A, Arras P, Zweibel E G
\ApJ{609}, 999 (2004)

\bibitem[Landau(1930)]{Landau30}
Landau L D
\artref{Z.\ f.\ Physik}{64}, 629 (1930)

\bibitem[Pavlov \& Gnedin(1984)]{PavlovGnedin}
Pavlov G G, Gnedin Yu N
\artref{Sov.\ Sci..\ Rev. E: Astrophys.\ Space Phys.}{3}, 197 (1984)

\bibitem[Ho \& Lai(2003)]{HoLai02}
Ho W C G, Lai D
\MNRAS{338}, 233 (2003)

\bibitem{Truemper78}
Tr\"umper J et al.
\ApJ{219}, L105 (1978)

\bibitem{Enoto}
Enoto T et al.
\artref{Publ.\ Astron.\ Soc.\ Pacific}{60}, S57 (2008)

\bibitem{RodesRoca}
Rodes-Roca J J et al.
\AandA{508}, 395 (2009)

\bibitem{Santangelo}
Santangelo A et al.
\ApJ{523}, L85 (1999)

\bibitem{Bignami}
Bignami G F et al.
\artref{Nature}{423}, 725 (2003)

\bibitem{SuPaW}
Suleimanov V F, Pavlov G G, Werner K
\ApJ{714}, 630 (2010)

\bibitem[Broderick \etal(2000)]{Broderick}
Broderick A, Prakash M, Lattimer J M
\ApJ{537}, 351 (2000)

\bibitem[Johnson \etal(1983)]{JHY83}
Johnson B R, Hirschfelder J O, Yang K H
\artref{Rev.\ Mod.\ Phys.}{55}, 109 (1983)

\bibitem[Ruder \etal(1994)]{Ruder}
Ruder H et al.
\textit{Atoms in Strong Magnetic Fields: Quantum Mechanical
Treatment and Applications in Astrophysics and Quantum Chaos}
(Berlin: Springer, 1994)

\bibitem[Ruderman(1971)]{Ruderman71}
Ruderman M A
\PRL{27}, 1306 (1971)

\bibitem[Potekhin(1998)]{P98}
Potekhin A Y
\artref{J.\ Phys. B}{31}, 49 (1998)

\bibitem[Detmer \etal(1998)Detmer, Schmelcher, \& Cederbaum]{Detmer}
Detmer T, Schmelcher P, Cederbaum L S
\artref{Phys.\ Rev. A}{57}, 1767 (1998)

\bibitem[Kappes \& Schmelcher(1996)]{KS96}
Kappes U, Schmelcher P
\artref{Phys.\ Rev. A}{53}, 3869 (1996)

\bibitem{Turbiner07}
Turbiner A V
\ApSS{308}, 267 (2007)

\bibitem[Medin \& Lai(2007)]{MedinLai06b}
Medin Z, Lai D
\artref{Phys.\ Rev. A}{74}, 062508 (2007)

\bibitem{Burkova}
Burkova et al,
\artref{Sov.\ Phys. JETP}{44}, 276 (1976)

\bibitem{IPM84}
Ipatova I P, Maslov A Yu, Subashiev A V 
      \artref{Sov. Phys. JETP}{60}, 1037 (1984)

\bibitem[Vincke \etal(1992)Vincke, Le Dourneuf, \& Baye]{VDB92}
Vincke M, Le Dourneuf M, Baye D
\artref{J.\ Phys.\ B}{25}, 2787 (1992)

\bibitem{P94}
Potekhin A Y
\artref{J.\ Phys.\ B}{27}, 1073 (1994)

\bibitem[Bezchastnov \etal(1998)Bezchastnov, Pavlov, \& Ventura]{BPV97}
Bezchastnov V G, Pavlov G G, Ventura J
\artref{Phys.\ Rev. A}{58}, 180 (1998)

\bibitem[Pavlov \& Bezchastnov(2005)]{PB05}
Pavlov G G, Bezchastnov V G
\ApJ{635}, L61 (2005)

\bibitem[Vincke \& Baye(1988)]{VB88}
Vincke M, Baye D
\artref{J.\ Phys. B}{21}, 2407 (1988)

\bibitem[Pavlov \& M\'esz\'aros(1993)]{PM93}
Pavlov G G, M\'esz\'aros P
\ApJ{416}, 752 (1993)

\bibitem[Yakovlev \& Kaminker(1994)]{YaK94}
Yakovlev D G, Kaminker A D,
in \textit{The Equation of State in Astrophysics}
(Eds G~Chabrier, E~Schatzman)
(Cambridge: Cambridge University Press, 1994), p.~214

\bibitem{P99}
Potekhin A Y
\AandA{351}, 787 (1999)

\bibitem[Ventura \& Potekhin(2001)]{VP01}
Ventura J, Potekhin A Y,
in \textit{The Neutron Star -- Black Hole Connection}
(Eds.\ C~Kouveliotou, E~P~J van den Heuvel, J~Ventura)
(Dordrecht: Kluwer, 2001) p.~393

\bibitem[Thorne(1977)]{Thorne77}
Thorne K S
\ApJ{212}, 825 (1977)

\bibitem[Gnedin \etal(2001)Gnedin, Yakovlev, \& Potekhin]{GYP}
Gnedin O Y, Yakovlev D G, Potekhin A Y
\MNRAS{324}, 725 (2001)

\bibitem[Potekhin \etal(2007)]{PCY07}
Potekhin A~Y, Chabrier G, Yakovlev D~G
\ApSS{353} (2007); corrected version
arXiv:astro-ph/0611014v3

\bibitem[Gudmundsson \etal(1983)]{GPE}
Gudmundsson E H, Pethick C J, Epstein R I
\ApJ{272}, 286 (1983)

\bibitem[Potekhin \etal(2003)]{refroid}
Potekhin A Y et al.
\ApJ{594}, 404 (2003)

\bibitem[Yakovlev \etal(2004)]{transie}
Yakovlev D G et al.
\AandA{417}, 169 (2004)

\bibitem{Yak08}
Yakovlev D G et al.
\artref{AIP Conf.\ Proc.}{983}, 379 (2008)

\bibitem[Chabrier \etal(1997)Chabrier, Potekhin, \& Yakovlev]{PCY97}
Potekhin A Y, Chabrier G, Yakovlev D G
\AandA{323}, 415 (1997)

\bibitem{CSP06}
Chabrier G, Saumon D, Potekhin A Y
\artref{J.\ Phys.\ A: Math.\ Gen.}{39}, 4411 (2006)

\bibitem{PUC}
Potekhin A Y, Urpin V A, Chabrier G
\AandA{443}, 1025 (2005)

\bibitem{PZ03}
Pavlov G G, Zavlin V E,
in \textit{Texas in Tuscany. XXI Texas Symposium on 
Relativistic Astrophysics}
(Eds. R Bandiera, R Maolino, F Manucci)
(Singapore: World Scientific, 2003)

\bibitem[Ho \etal(2007)]{Ho_ea07}
Ho W C G et al.
\MNRAS{375}, 821 (2007)

\bibitem{Mihalas}
Mihalas D
\textit{Stellar Atmospheres}
(2nd ed.)
(San Francicso: Freeman, 1978)

\bibitem{Walter97}
Walter F M, Matthews L D
\artref{Nature}{389}, 358 (1997)

\bibitem{Walter01}
Walter F M
\ApJ{549}, 433 (2001)

\bibitem{Drake}
Drake J J et al.
\ApJ{572}, 996 (2002)

\bibitem{WL02}
Walter F M, Lattimer J M
\ApJ{576}, L145 (2002)

\bibitem{Kaplan02}
Kaplan D L, van Kerkwijk M H, Anderson J
\ApJ{571}, 447 (2002)

\bibitem{vKK07}
van Kerkwijk M H, Kaplan D L
\ApSS{308}, 191 (2007)

\bibitem{Pons}
Pons J A et al.
\ApJ{564}, 981 (2002)

\bibitem{Burwitz}
Burwitz V et al.
\AandA{399}, 1109 (2003)

\bibitem{Walter10}
Walter F M et al.
\ApJ{724}, 669 (2010)

\bibitem{Demorest}
Demorest P et al.
Nature \textbf{467} 1081 (2010)

\end{thebibliography}
\end{document}